\documentclass[a4paper,12pt]{article}
\usepackage{amssymb,graphicx,float,epstopdf,amsmath,amsfonts,amsthm,cases,cite,epsfig,subfigure}
\usepackage[symbol*]{footmisc}
\graphicspath{{pics/}}
\DeclareMathOperator\artanh{artanh}
\usepackage[pdfencoding=auto]{hyperref}
\usepackage{xcolor}
\usepackage{hyperref}
\definecolor{linkcolor}{HTML}{799B03}
\definecolor{urlcolor}{HTML}{799B03}
\hypersetup{pdfstartview=FitH,linkcolor=linkcolor,urlcolor=urlcolor,colorlinks=true}
\begin{document}	
\begin{flushright}
INR-TH-2017-015
\end{flushright}

\vspace{10pt}

\begin{center}
{\LARGE \bf Cosmological bounce and Genesis}

\vspace{0.4cm}

{\LARGE \bf beyond Horndeski.}

\vspace{20pt}

R. Kolevatov$^{a,b}$\footnote[1]{\textbf{e-mail:} kolevatov@ms2.inr.ac.ru},
S. Mironov$^{a,c}$\footnote[2]{\textbf{e-mail:} sa.mironov\_1@physics.msu.ru},
N. Sukhov$^{b}$\footnote[3]{\textbf{e-mail:} nd.sukhov@physics.msu.ru},
V. Volkova$^{a,b}$\footnote[4]{\textbf{e-mail:} volkova.viktoriya@physics.msu.ru}

\vspace{15pt}

$^a$\textit{Institute for Nuclear Research of the Russian Academy of Sciences,\\
60th October Anniversary Prospect, 7a, 117312 Moscow, Russia}\\

$^b$\textit{Department of Particle Physics and Cosmology, Physics Faculty,\\
M.V. Lomonosov Moscow State University,\\
Vorobjevy Gory, 119991 Moscow, Russia}
	
$^c$\textit{Institute for Theoretical and Experimental Physics\\
Bolshaya Cheriomyshkinskaya, 25, 117218 Moscow, Russia}
\end{center}

\vspace{5pt}

\begin{abstract}
We study ``classical'' bouncing and Genesis models in beyond Horndeski theory.
We give an example of spatially flat bouncing solution that is non-singular and stable
throughout the whole evolution. We also provide an example of stable geodesically
complete Genesis with similar features. The model is arranged in such a way that the
scalar field driving the cosmological evolution initially behaves like full-fledged beyond
Horndeski, whereas at late times it becomes a  massless scalar field minimally coupled
to gravity.
\end{abstract}

\section{Introduction}
Bouncing cosmologies are interesting scenarios that are complementary or alternative to inflation. These models enable one to solve the singularity problem, and at the same time possess  many nice features of inflationary theories. 

A model of spatially flat bounce requires the violation of the Null Energy Condition
(NEC), which may be realised by some fairly unconventional matter.  One of the possible
candidates for the latter are the generalized Galileons, that are equivalent to Horndeski
theory~\cite{Horndeski:1974wa, Fairlie:1991qe, Luty:2003vm, Nicolis:2008in, Deffayet:2010zh, Deffayet:2010qz, Kobayashi:2011nu, Padilla:2012dx}. Galileons are scalar fields, whose Lagrangians involve second derivatives but
the corresponding field equations are still of the second order (for a review see, e.g.~\cite{Rubakov:2014jja}). 

It was shown, that generalized Galileons can indeed be utilized for constructing bouncing
Universe models, which have no pathologies (ghosts, gradient instabilities, etc.) around
the moment of bounce~\cite{Qiu:2011cy, Easson:2011zy, Osipov:2013ssa, Qiu:2013eoa}. However, the issue of avoiding instabilities becomes more
challenging if one considers more complete models. Indeed, with generalized Galileons,
when trying to extend pathology-free bounce solutions to long enough periods of time, one
always meets gradient (or, possibly, ghost) instabilities~\cite{Cai:2012va,  Koehn:2013upa, Battarra:2014tga, Qiu:2015nha, Kobayashi:2015gga, Wan:2015hya, Ijjas:2016tpn}. Similar observations apply
to (``complete'') Genesis models~\cite{Kobayashi:2015gga,Pirtskhalava:2014esa}.

One of the possible ways to get around the gradient instability problem is to arrange a
model in such a way that the term in the action, which is quadratic in spatial gradient and
has the wrong sign, is small so that higher derivative terms restore stability at sufficiently
high momenta~\cite{Creminelli:2006xe, Pirtskhalava:2014esa}. Another possibility is that the strong coupling momentum scale is
low enough~\cite{Koehn:2015vvy}. In both cases the exponential growth of perturbations does not lead to
a catastrophic outcome, provided that the cosmic time interval when the instabilities are
present, is short.

On the other hand, it is  definitely of interest whether geodesically complete, healthy
cosmological models --- the flat bouncing scenario and Genesis in particular --- can be
constructed on the basis of the generalized Galileons. The inevitable presence of pathologies
within a subclass of Horndeski theories was formulated as a no-go theorem~\cite{Libanov:2016kfc} and further
generalised to the case of direct interaction of the Galileon field with conventional scalar
field~\cite{Kolevatov:2016ppi}. The further extension of the no-go argument was presented in~\cite{Kobayashi:2016xpl}, showing that
complete cosmological models, based on the most general Horndeski theories, are always
plagued with instabilities. This no-go theorem has been further generalized to a multi
Galileon case~\cite{Akama:2017jsa}. There was a loophole, though, mentioned in~\cite{Kobayashi:2016xpl} in the Genesis context:
it is possible to construct  non-pathological bouncing or Genesis solution, in which strong
coupling gravity regime occurs long before or after the bouncing stage~\cite{Ijjas:2016vtq}. 

Recently, it has been proposed that Horndeski theories can be safely extended~\cite{Gleyzes:2014dya}. Even
though the newly introduced Lagrangian terms lead in general  to the presence of third
derivatives in the equations of motion, the theory still does not contain extra degrees of
freedom. These new terms in the Lagrangian may hopefully be utilized for solving the
problem of instabilities in the complete cosmologies. This expectation has been strongly
substantiated within the Effective Field Theory (EFT) approach~\cite{Creminelli:2016zwa, Cai:2016thi}, see Sec.~\ref{sec:generalities} for further discussion. 

Our purpose in this paper is to present the full Lagrangians of the models, belonging to
the class of beyond Horndeski theory, and construct classical bouncing and Genesis solutions
which are non-singular during the entire evolution. Our set up does not have any kind of
fine-tuning, nor it gets into  strongly coupled gravity regime. The gradient and/or ghost
instabilities are absent about our backgrounds at any time. We also address the issue of
scale covariance at early times, which is behind the simple form of  our early-time solution.
In the Genesis case, we explicitly show the geodesic completeness of our solution.

The paper is organized as follows. In Sec.~\ref{sec:generalities} we give the general Lagrangian of beyond
Horndeski theory and obtain the quadratic action for perturbations in the unitary gauge. We
provide a recipe for getting around the no-go argument existing in the generalized Galileon
theories. We construct a healthy bouncing solution in Sec.~\ref{sec:bounce} and give an analysis of its
stability. There we also discuss the issue of early-time scale covariance of the theory. In
Sec.~\ref{sec:genesis} an example of healthy and geodesically complete Genesis is described. We conclude
in Sec.~\ref{sec:conclusion}.

{\bf Note added:} while this paper was close to completion, a preprint~\cite{Cai:2017dyi} appeared with
different construction for stable classical bounce in beyond Horndeski theory.

\section{Stability of solutions in beyond Horndeski model}
\label{sec:generalities}
We consider beyond Horndeski theory with the following action: 
\begin{subequations}
\label{eq:action_setup}
\begin{align}
&S=\int\mathrm{d}^4x\sqrt{-g}\left(\mathcal{L}_2 + \mathcal{L}_3 + \mathcal{L}_4 + \mathcal{L}_5 + \mathcal{L_{BH}}\right),\\
&\mathcal{L}_2=F(\pi,X),\\
&\mathcal{L}_3=K(\pi,X)\Box\pi,\\
&\mathcal{L}_4=-G_4(\pi,X)R+2G_{4X}(\pi,X)\left[\left(\Box\pi\right)^2-\pi_{;\mu\nu}\pi^{;\mu\nu}\right],\\
&\mathcal{L}_5=G_5(\pi,X)G^{\mu\nu}\pi_{;\mu\nu}+\frac{1}{3}G_{5X}\left[\left(\Box\pi\right)^3-3\Box\pi\pi_{;\mu\nu}\pi^{;\mu\nu}+2\pi_{;\mu\nu}\pi^{;\mu\rho}\pi_{;\rho}^{\;\;\nu}\right],\\
&\mathcal{L_{BH}}=F_4(\pi,X)\epsilon^{\mu\nu\rho}_{\quad\;\sigma}\epsilon^{\mu'\nu'\rho'\sigma}\pi_{,\mu}\pi_{,\mu'}\pi_{;\nu\nu'}\pi_{;\rho\rho'}\\
\nonumber&\qquad+F_5(\pi,X)\epsilon^{\mu\nu\rho\sigma}\epsilon^{\mu'\nu'\rho'\sigma'}\pi_{,\mu}\pi_{,\mu'}\pi_{;\nu\nu'}\pi_{;\rho\rho'}\pi_{;\sigma\sigma'},
\end{align}
\end{subequations}
where $\pi$ is the Galileon field,  $\pi_{,\mu}=\partial_\mu\pi$, $X=g^{\mu\nu}\pi_{,\mu}\pi_{,\nu}$, $\pi_{;\mu\nu}=\triangledown_\nu\triangledown_\mu\pi$, $\Box\pi = g^{\mu\nu}\triangledown_\nu\triangledown_\mu\pi$,
$G_{4X}=\partial G_4/\partial X$, etc. Horndeski theory is obtained if one sets $F_4(\pi,X)=F_5(\pi,X)=0$. It
is worth noting that action~\eqref{eq:action_setup} already has gravitational parts, furthermore, one can restore
the Einstein-Hilbert gravity by choosing $G_4(\pi,X)=\frac{1}{2\kappa}$, $G_5(\pi,X)=0$, where $\kappa = 8\pi G$ and
$G$ is the gravitational constant. We start from the unperturbed spatially flat FLRW metric
(mostly negative signature)
\begin{equation*}
\mathrm{d}s^2 = \mathrm{d}t^2 - a^2(t)\delta_{ij}\mathrm{d}x^i\mathrm{d}x^j.
\end{equation*}
Variation of action~\eqref{eq:action_setup} with respect to $g^{00}$ and $g^{ii}$ leads to the equations of motion
\begin{subequations}
\label{eq:einst_eq_setup}
\begin{align}
\delta g^{00}:\quad
&F-2F_XX-6HK_XX\dot{\pi}+K_{\pi}X+6H^2G_4+6HG_{4\pi}\dot{\pi}
\\\nonumber&-24H^2X(G_{4X}+G_{4XX}X)+12HG_{4\pi X}X\dot{\pi}
\\\nonumber&-2H^3X\dot{\pi}(5G_{5X}+2G_{5XX}X)+3H^2X(3G_{5\pi}+2G_{5\pi X}X)
\\\nonumber&+6H^2X^2(5F_4+2F_{4X}X)+6H^3X^2\dot{\pi}(7F_5+2F_{5X}X)=0,\\
\delta g^{ii}:\quad
&F-X(2K_X\ddot{\pi}+K_\pi)+2(3H^2+2\dot{H})G_4-12H^2G_{4X}X
\\\nonumber&-8\dot{H}G_{4X}X-8HG_{4X}\ddot{\pi}\dot{\pi}-16HG_{4XX}X\ddot{\pi}\dot{\pi}
\\\nonumber&+2(\ddot{\pi}+2H\dot{\pi})G_{4\pi}+4XG_{4\pi X}(\ddot{\pi}-2H\dot{\pi})+2XG_{4\pi\pi}
\\\nonumber&-2XG_{5X}(2H^3\dot{\pi}+2H\dot{H}\dot{\pi}+3H^2\ddot{\pi})-4H^2G_{5XX}X^2\ddot{\pi}
\\\nonumber&+G_{5\pi}(3H^2X+2\dot{H}X+4H\ddot{\pi}\dot{\pi})+2HG_{5\pi X}X(2\ddot{\pi}\dot{\pi}-HX)
\\\nonumber&+2HG_{5\pi\pi}X\dot{\pi}+2F_4X(3H^2X+2\dot{H}X+8H\ddot{\pi}\dot{\pi})+8HF_{4X}X^2\ddot{\pi}\dot{\pi}
\\\nonumber&+4HF_{4\pi}X^2\dot{\pi}+6HF_5X^2(2H^2\dot{\pi}+2\dot{H}\dot{\pi}+5H\ddot{\pi})
\\\nonumber&+12H^2F_{5X}X^3\ddot{\pi}+6H^2F_{5\pi}X^3=0,
\end{align}
\end{subequations}
where dot denotes the derivative with respect to $t$ and $H=\dot{a}/a$ is the Hubble parameter.

We study  Lagrangian for perturbations in the unitary gauge, which means that we choose
perturbations of the scalar field equal to   zero, $\delta\pi=0$. We adopt the following 
parametrization:
\begin{equation*}
g_{00}=1+2\alpha,\quad g_{0i}=-\partial_i\beta,\quad g_{ij}=-a^2\left(e^{2\zeta}\delta_{ij}+h_{ij}^T\right),
\end{equation*}
where $h_{ij}^T$ denotes tensor perturbations: $h^T_{ii}=0,\quad\partial_ih^T_{ij}=0$. Our concern is the stability
against the most rapidly developing perturbations, which is determined by the UV behavior
of the theory. Therefore, we only keep terms with the highest order of spatial and temporal
derivatives and the Lagrange multipliers. Quadratic action for perturbations then reads:
\begin{equation}
\label{eq:pert_action_setup}
\begin{aligned}
S=\int\mathrm{d}t\mathrm{d}^3x a^3\Bigg[\left(\dfrac{\mathcal{\hat{G}_T}}{8}\left(\dot{h}^T_{ik}\right)^2-\dfrac{\mathcal{F_T}}{8a^2}\left(\partial_i h_{kl}^T\right)^2\right)+\Big(-3\mathcal{\hat{G}_T}\dot{\zeta}^2+\mathcal{F_T}\dfrac{(\triangledown\zeta)^2}{a^2}\\
-2\mathcal{G_T}\alpha\dfrac{\triangle\zeta}{a^2}+2\mathcal{\hat{G}_T}\dot{\zeta}\dfrac{\triangle\beta}{a^2}+6\Theta\alpha\dot{\zeta}-2\Theta\alpha\dfrac{\triangle\beta}{a^2}+\Sigma\alpha^2\Big)\Bigg],
\end{aligned}
\end{equation}
with the following coefficients:
\begin{subequations}
\label{eq:Full_coeff}
\begin{align}
\label{eq:GT_coeff_setup}
&\mathcal{G_T}=2G_4-4G_{4X}X+G_{5\pi}X-2HG_{5X}X\dot{\pi},\\
&\mathcal{F_T}=2G_4-2G_{5X}X\ddot{\pi}-G_{5\pi}X,\\
\label{eq:D_coeff_setup}
&\mathcal{D}=2F_4X\dot{\pi}+6HF_5X^2,\\
&\mathcal{\hat{G}_T}=\mathcal{G_T}+\mathcal{D}\dot{\pi},\\
\label{eq:Theta_coeff_setup}
&\Theta=-K_XX\dot{\pi}+2G_4H-8HG_{4X}X-8HG_{4XX}X^2+G_{4\pi}\dot{\pi}+2G_{4\pi X}X\dot{\pi}\\
\nonumber&-5H^2G_{5X}X\dot{\pi}-2H^2G_{5XX}X^2\dot{\pi}+3HG_{5\pi}X+2HG_{5\pi X}X^2\\
\nonumber&+10HF_4X^2+4HF_{4X}X^3+21H^2F_5X^2\dot{\pi}+6H^2F_{5X}X^3\dot{\pi},\\
\label{eq:Sigma_coeff_setup}
&\Sigma=F_XX+2F_{XX}X^2+12HK_XX\dot{\pi}+6HK_{XX}X^2\dot{\pi}-K_{\pi}X-K_{\pi X}X^2\\
\nonumber&-6H^2G_4+42H^2G_{4X}X+96H^2G_{4XX}X^2+24H^2G_{4XXX}X^3\\
\nonumber&-6HG_{4\pi}\dot{\pi}-30HG_{4\pi X}X\dot{\pi}-12HG_{4\pi XX}X^2\dot{\pi}+30H^3G_{5X}X\dot{\pi}\\
\nonumber&+26H^3G_{5XX}X^2\dot{\pi}+4H^3G_{5XXX}X^3\dot{\pi}-18H^2G_{5\pi}X-27H^2G_{5\pi X}X^2\\
\nonumber&-6H^2G_{5\pi XX}X^3-90H^2F_4X^2-78H^2F_{4X}X^3-12H^2F_{4XX}X^4\\
\nonumber&-168H^3F_5X^2\dot{\pi}-102H^3F_{5X}X^3\dot{\pi}-12H^3F_{5XX}X^4\dot{\pi}.
\end{align}
\end{subequations}
The key property specific to beyond Horndeski theory as compared to Horndeski is the
difference between $\mathcal{\hat{G}_T}$ and $\mathcal{G_T}$, which comes precisely from beyond Horndeski term $\dot{\pi}\mathcal{D}$. This
new term will allow us to construct a stable beyond Horndeski bounce. 

Variation of action~\eqref{eq:pert_action_setup} with respect to the Lagrange multipliers leads to the constraint
equations: 
\begin{subequations}
\begin{align}
\label{eq:constr_setup}
\dfrac{\triangle\beta}{a^2}:&\qquad \alpha=\dfrac{\mathcal{\hat{G}_T}\dot{\zeta}}{\Theta},\\
\alpha:&\qquad\dfrac{\triangle\beta}{a^2}=\dfrac{1}{\Theta}\left(3\Theta\dot{\zeta}-\mathcal{G_T}\dfrac{\triangle\zeta}{a^2}+\Sigma\alpha\right).
\end{align}
\end{subequations}
Substituting these into action~\eqref{eq:pert_action_setup} we obtain quadratic action in the unconstrained form
(similar expression has been obtained in ADM formalism in Ref.~\cite{Gleyzes:2014dya})
\begin{equation}
\label{eq:unconstrained_action}
S=\int\mathrm{d}t\mathrm{d}^3x a^3\left[\dfrac{\mathcal{\hat{G}_T}}{8}\left(\dot{h}^T_{ik}\right)^2-\dfrac{\mathcal{F_T}}{8a^2}\left(\partial_i h_{kl}^T\right)^2+\mathcal{G_S}\dot{\zeta}^2-\mathcal{F_S}\dfrac{(\triangledown\zeta)^2}{a^2}\right],\\
\end{equation}
where the coefficients are
\begin{subequations}
\label{eq:Scal_setup}
\begin{align}
&\mathcal{G_S}=\dfrac{\Sigma\mathcal{\hat{G}_T}^2}{\Theta^2}+3\mathcal{\hat{G}_T},\\
\label{eq:FS_setup}
&\mathcal{F_S}=\dfrac{1}{a}\dfrac{\mathrm{d}\xi}{\mathrm{d}t}-\mathcal{F_T},\\
\label{eq:xi_func_setup}
&\xi=\dfrac{a\mathcal{G_T}\mathcal{\hat{G}_T}}{\Theta}=\dfrac{a\left(\mathcal{\hat{G}_T}-\mathcal{D}\dot{\pi}\right)\mathcal{\hat{G}_T}}{\Theta}.
\end{align}
\end{subequations}
The speeds of sound for tensor and scalar perturbations are, respectively,
\begin{equation*}
c_\mathcal{T}^2=\dfrac{\mathcal{F_T}}{\mathcal{\hat{G}_T}},\qquad c_\mathcal{S}^2=\dfrac{\mathcal{F_S}}{\mathcal{G_S}}.
\end{equation*}
A healthy and stable solution requires correct signs for kinetic and gradient terms as well as
subluminal propagation:
\begin{equation}
\label{eq:stability_cond_setup}
\mathcal{\hat{G}_T} > \mathcal{F_T}>0,\quad \mathcal{G_S}>\mathcal{F_S}>0.
\end{equation}

Horndeski theories are known to prohibit the stable bounce. Let us briefly summarize the
proof of the corresponding no-go theorem and show how to bypass it by introducing  beyond
Horndeski terms. We will largely follow the proof given in Refs.~\cite{Libanov:2016kfc},~\cite{Kobayashi:2016xpl}. The proof starts
with Eq.~\eqref{eq:FS_setup} expressed in an integrated form:
\begin{equation}
\label{eq:Integ_setup}
\xi(t_2)-\xi(t_1)=\int\limits_{t_1}^{t_2}a(t)\left(\mathcal{F_T}+\mathcal{F_S}\right)\mathrm{d}t.
\end{equation}  
	
For a bounce, the scale factor initially decreases, reaches minimum and then starts
increasing. Thus, it is bounded by its minimal value at the moment of bounce  $a(t)\geq a_{min}>0$.
Furthermore, we want to keep $\mathcal{F_T}$ and $\mathcal{F_S}$ positive and bounded from below, otherwise we
are back to the option of the domination of higher order spatial derivatives~\cite{Creminelli:2006xe}. This means
that the combination $a(t)(\mathcal{F_T}+\mathcal{F_S})$ is always positive and bounded from below. The same
argument applies to geodesically complete bounce~\cite{Creminelli:2016zwa}.

Now, suppose that $\xi(t_2)>0$. We have
\begin{equation*}
\xi(t_1)=\xi(t_2) - \int\limits_{t_1}^{t_2}a(t)\left(\mathcal{F_T}+\mathcal{F_S}\right)\mathrm{d}t \; .
\end{equation*}
This shows that at early enough times $t_1$, one has $\xi(t_1)<0$. Another possibility is that
$\xi(t_1)<0$. Then we write
\begin{equation*}
\xi(t_2)= - |\xi(t_1)| + \int\limits_{t_1}^{t_2}a(t)\left(\mathcal{F_T}+\mathcal{F_S}\right)\mathrm{d}t,
\end{equation*}
and at large enough $t_2$ one has $\xi(t_2)>0$.

Hence, there must be a moment of time when $\xi(t)$ changes sign, i.e, it crosses zero,
$\xi(t_0)=0$. This property is valid both in Horndeski and beyond Horndeski theories. In the
Horndeski case $\mathcal{\hat{G}_T}=\mathcal{G_T}$, and, therefore, Eq.~\eqref{eq:xi_func_setup} reads $\xi=\frac{a\mathcal{G_T}^2}{\Theta}$. Hence, zero crossing for $\xi$
requires $\Theta\rightarrow \infty$, which means a singularity of Lagrangian functions in action~\eqref{eq:action_setup}.\footnote{Another option is to let $\mathcal{G_T}$ and $\Theta$ become zero at the time when $\xi=0$ ~\cite{Ijjas:2016vtq}. At least naively, the theory becomes strongly coupled near $\xi=0$ in this case. Yet another possibility is that  $\mathcal{F_T}\rightarrow 0$ and $\mathcal{F_S}\rightarrow 0$ as $t\rightarrow -\infty$~\cite{Kobayashi:2016xpl,Ijjas:2016vtq}, so that the integral~\eqref{eq:Integ_setup} is convergent in the lower limit. This again brings one to a strong gravity regime.}

In contrast to Horndeski theory, the expression for $\xi$ beyond Horndeski involves $(\mathcal{\hat{G}_T} \cdot \mathcal{G_T})$,
rather than $\mathcal{{G}_T}^2$. $\mathcal{G_T}$ is no longer a physically important kinetic coefficient, so it can cross
zero and be negative at large negative times. At the same time, with the extra term $\dot{\pi}D$ in~\eqref{eq:D_coeff_setup} it is possible to have $\mathcal{\hat{G}_T}$ positive at all times. Hence, the difference between $\mathcal{G_T}$ and $\mathcal{\hat{G}_T}$
enables one to have regular $\xi=0$ point with positive $\mathcal{\hat{G}_T}$ everywhere. This can be achieved
only as a joint play of $\mathcal{L}_4$ or $\mathcal{L}_5$ Lagrangians together with $\mathcal{L_{BH}}$, see eqs.~\eqref{eq:GT_coeff_setup} and~\eqref{eq:D_coeff_setup}.

We note that the same arguments apply to the Genesis case. Namely, healthy Genesis
is impossible without introducing beyond Horndeski terms (modulo ``loopholes'' related to
strong gravity regime and geodesic incompleteness~\cite{Libanov:2016kfc, Kobayashi:2016xpl}). On the other hand, the difference
between $\mathcal{\hat{G}_T}$ and $\mathcal{G_T}$ suggests that healthy Genesis may be possible in beyond Horndeski
theory.

The quadratic action for perturbations in beyond Horndeski theory was also obtained
within the EFT approach~\cite{Cai:2016thi, Creminelli:2016zwa}, where a similar set of  coefficients was found. The
correspondence between the coefficients in eqs.~\eqref{eq:Full_coeff} and~\eqref{eq:Scal_setup} and ones given in Ref.~\cite{Creminelli:2016zwa} is
\begin{equation*}
\begin{aligned}[c]
&\mathcal{\hat{G}_T}=M_{Pl}^2, \\
&\mathcal{G_T}=M_{Pl}^2-2\bar{m}_3^2, \\
&\Sigma=\frac{1}{2}(m_2^4-2M_{Pl}^2\dot{H}-6M_{Pl}^2H^2),
\end{aligned}
\qquad
\begin{aligned}[c]
&\mathcal{F_T}=M_{Pl}^2, \\
&\xi =Y, \\
&\Theta=M_{Pl}^2H-\frac{\hat{m}_1^3}{2}.
\end{aligned}
\end{equation*}
Similar relations can be obtained for Ref.~\cite{Cai:2016thi}. 

Refs.~\cite{Cai:2016thi} and~\cite{Creminelli:2016zwa} proposed the same idea to evade the no-go argument as we have
described above. Namely, the beyond Horndeski term $\bar{m}_3^2$ (which equals to $\dot{\pi}\mathcal{D}$ in our language)
was used to allow $Y$ ($\xi$ in our case) to safely cross zero, while ensuring positive kinetic
coefficient $M_{Pl}^2$ ($\mathcal{\hat{G}_T}$ in our case). However, the EFT approach does not enable one to write the
corresponding 4D Lagrangian~\eqref{eq:action_setup} as well as to check whether the equations of motion~\eqref{eq:einst_eq_setup}
are satisfied. 

Before we jump to a specific example of a healthy beyond Horndeski bounce, there is an
issue to discuss. Let us consider the case of $\Theta$ crossing zero at some moment $t_0$. In order to
have regular $\mathcal{F_S}$ and $\mathcal{G_S}$ one would have to fine tune the coefficients $\Sigma$ and $\mathcal{G_T}$ (see eqs.~\eqref{eq:Scal_setup}).
The numerators in $\mathcal{F_S}$ and $\mathcal{G_S}$ have to have higher order zeros at the same moment of time
$t_0$. It is this case that was considered when constructing the example for a complete bounce in Ref.~\cite{Creminelli:2016zwa}. 
	
We are going to avoid this fine tuning and keep $\Theta\neq0$ at all times. This comes with
a price: the Galileon field $\pi$ cannot be conventional scalar field, described solely by the
Lagrangian $\mathcal{L}_2$ and the Einstein--Hilbert term, in {\itshape both} distant past and future. Otherwise,
Eq.~\eqref{eq:Theta_coeff_setup} states that such conventional scalar field has $\Theta$ proportional to the Hubble parameter,
$\Theta=2G_4H$, where $G_4$ is the gravitational constant; the bounce starts
with $H<0$ and ends with $H>0$, therefore, $\Theta$ must change sign and cross zero at some point. On the contrary,
in this paper we are going to construct the bouncing scenario with non-trivial Galileon field
$\pi$ in the asymptotic past, which eventually evolves into a conventional scalar field in distant
future. In this way we keep $\Theta$ finite and positive at all times.

It is impossible also to cook up a Genesis model where the Galileon is massless scalar
field both at late and early times. The argument is slightly more involved as compared to
the bounce case. It goes as follows. We want to have Einstein--Hilbert gravity at distant
past, so we impose at early times
\begin{equation*}
G_4(\pi,X) = \dfrac{1}{2\kappa},\quad G_5(\pi,X) = 0.
\end{equation*}
Thus, at early times one has
\begin{equation}
\mathcal{G_T}=2G_4-4G_{4X}X+G_{5\pi}X-2HG_{5X}X\dot{\pi} = \dfrac{1}{\kappa}.
\label{jun2-17-10}
\end{equation}
However, according to the mechanism of avoiding the no-go theorem, $\mathcal{G_T}$ must change sign
at some point and is negative at early times, in  contradiction with~\eqref{jun2-17-10}. There is a possible
loophole: the latter argument assumes that $\mathcal{G_T}$ crosses zero only once. To see that this is
the case we recall that
\begin{equation*}
\dfrac{\mathrm{d}\xi}{\mathrm{d}t} > a\left(\mathcal{F_T} + \mathcal{F_S}\right),
\end{equation*}
so the function $\xi$ is always growing. Therefore, it can cross zero only once. According to our
discussion above, $\mathcal{G_T}$ crosses zero at the same moment as $\xi$, thus $\mathcal{G_T}$ crosses zero only once
as well. This completes the argument.

Note that the same argument works in the case of bounce. It does not rely upon the behavior of $\Theta$ and irrespectively of this
behavior shows that it is impossible to cook up a
bouncing model where the Galileon is massless scalar field both at late and early times.

\section{Cosmological bounce: an example}
\label{sec:bounce}
In this section we give a concrete example of the Lagrangian functions and solution to the
field equations which describes a classical, fully stable cosmological bounce. We perform the
analysis in Planck units and set
\begin{equation*}
\kappa = 8\pi G = 1.
\end{equation*}
Of course, upon introducing an appropriate dimensionless parameter in the Lagrangian
functions, the resulting dynamics can be made safely sub-Planckian.

\begin{figure}[H]
\center{\includegraphics[width=0.7\linewidth]{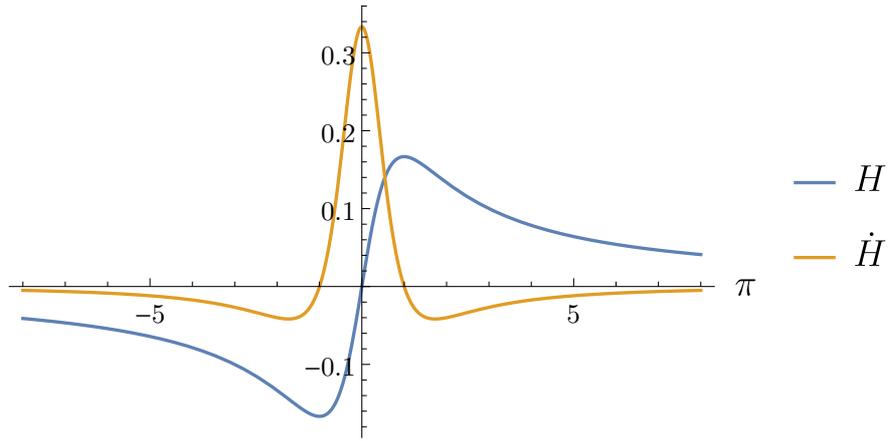} }
\caption{Hubble parameter $H$ and its time derivative $\dot{H}$.}
\label{pic:hubble}
\end{figure}

Before we choose an explicit form of the Lagrangian functions, let us recall the basic
requirement in our scenario. At late times, we require the Galileon to become a conventional
massless scalar field, whose equation of state is $p = \rho$, where $p$ is the pressure and $\rho$ is the
energy density.  Then, the late-time asymptotic of the Hubble parameter is
\begin{equation*}
t \rightarrow +\infty:\quad H(t) = \dfrac{1}{3t}.
\end{equation*}
Our purpose is to find the Lagrangian functions such that the model admits healthy bounce
solution. To this end, we choose the time dependence of {\it some} of the background functions
at our will, and reconstruct other functions by solving the field equations (so, we adopt
a general approach advocated in Refs.~\cite{Libanov:2016kfc,Kobayashi:2016xpl,Ijjas:2016tpn}). In particular, we choose the Hubble
parameter at all times equal to
\begin{equation}
\label{eq:hubble_bounce}
H(t) = \dfrac{t}{3(1 + t^2)},
\end{equation}
so the bounce occurs at $t = 0$. For the scale factor $a(t)$ one has
\begin{equation}
\label{eq:scale_factor_bounce}
a(t) = (1 + t^2)^\frac{1}{6}.
\end{equation}
The evolution of the Hubble parameter and its time derivative are shown in Fig.~\ref{pic:hubble}.
	
Let us choose the Lagrangian functions as follows:
\begin{subequations}
\label{eq:gal_functions_bounce}
\begin{align}
&F(\pi,X) = f_0(\pi) + f_1(\pi)X + f_2(\pi)X^2,\\
&K(\pi,X) = k_1(\pi)X,\\
&G_4(\pi,X) = \dfrac{1}{2} + g_{40}(\pi) + g_{41}(\pi)X,\\
&G_5(\pi,X) = 0,\\
&F_4(\pi,X) = f_{40}(\pi),\\
&F_5(\pi,X) = 0.
\end{align}
\end{subequations}
Note that it is sufficient for our purposes to set $G_5 = F_5 = 0$. We make a field redefinition
in such a way that that the background scalar field rolls with constant speed:
\begin{equation}
\label{eq:pi_bounce}
\pi = t.
\end{equation}
We immediately obtain that $X = \left(\partial\pi\right)^2 = 1$. For derivatives of the Lagrangian functions
one has
\begin{equation*}
\begin{aligned}
&F_X = f_1(t) + 2f_2(t),\quad F_{XX} = 2f_2(t),\\
&K_X = k_1(t),\\
&G_{4X} = g_{41}(t).
\end{aligned}
\end{equation*}
Our implementation of the approach outlined above is as follows: we first specify the explicit
forms of $k_1(t)$, $g_{40}(t)$, $g_{41}(t)$ and $f_{40}(t)$ at all times. Then we express the functions $f_1(t)$ and
$f_2(t)$ through $k_1(t)$, $g_{40}(t)$, $g_{41}(t)$ and $f_{40}(t)$ via the Einstein equations, again at all times.
The most tricky thing is then to specify the remaining function $f_0(t)$ in such a way that the
conditions of ghost absence and subluminality are satisfied. 
	
We now turn to functions $k_1(t)$, $g_{40}(t)$, $g_{41}(t)$ and $f_{40}(t)$. Since at late times, the resulting
Lagrangian should correspond to a theory of free massless scalar field and Einstein-Hilbert
gravity, it follows from Eq.~\eqref{eq:action_setup} that we should have the following late-time asymptotics of $K$,
$G_4$ and $F_4$:
\begin{equation}
\label{eq:K_G4_F4_late_times_bounce}
t \rightarrow +\infty:\quad K(\pi,X) = 0,\quad G_4(\pi,X) = \dfrac{1}{2},\quad F_4(\pi,X) = 0.
\end{equation}
On the other hand, as we discussed in the previous section, it is impossible to cook up a
scenario with a massless scalar field both at late and early times. So, the early-time behavior
of our system is less constrained. One of the possible (and pretty arbitrary) options for early-
time asymptotics of $G_4$ and $F_4$ is
\begin{equation}
\label{eq:G4_F4_early_times_bounce}
t \rightarrow -\infty:\quad G_4(\pi,X) = const,\quad F_4(\pi,X) = const.
\end{equation}
In contrast, a convenient choice for $K$ is that it is {\it not} equal to constant at early times. Since
we have defined asymptotics of $G_4$, $F_4$ and Hubble parameter, it is natural to conjecture the
behavior of $K$ by considering expression for $\Theta$, because it includes all these functions and
does not include $F$. From~\eqref{eq:Theta_coeff_setup} and~\eqref{eq:gal_functions_bounce} one has
\begin{equation*}
\Theta = - k_1 + \dot{g}_{40} + 3\dot{g}_{41} - 8g_{41}H + 2\left(\dfrac{1}{2} + g_{40} + g_{41}\right)H + 10f_{40}H.
\end{equation*}
Considering the latter equation at early times and taking into account Eq.~\eqref{eq:G4_F4_early_times_bounce}, one  concludes
that $k_1(t)$ is naturally chosen proportional to $H(t)$ to have a simple power-law behavior of
$\Theta$.

Taking into account asymptotics of Lagrangian functions~\eqref{eq:K_G4_F4_late_times_bounce},~\eqref{eq:G4_F4_early_times_bounce} and the fact that $K$
should be proportional to Hubble parameter at early times, let us choose $k_1(t)$, $g_{40}(t)$, $g_{41}(t)$
and $f_{40}(t)$ in the following form
\begin{subequations}
\label{eq:k1_g40_g41_f40_bounce}
\begin{align}
&k_1(t) = c_1\dfrac{t}{1+t^2}\left(1-\tanh (t)\right),
\label{jun2-17-1}\\
&g_{40}(t) = c_2\left(1-\tanh(t)\right),\quad g_{41}(t) = c_3\left(1-\tanh(t)\right),
\label{jun2-17-2}\\
&f_{40}(t) = c_4\left(1-\tanh(t)\right),
\label{jun2-17-3}
\end{align}
\end{subequations}
where $c_1$, $c_2$, $c_3$ and $c_4$ are constant coefficients.
	
Our main purpose now is to find time dependence of $f_0(t)$, $f_1(t)$ and $f_2(t)$. Note that
the only non-vanishing term in the Lagrangian at late times is $f_1(t)$, since it corresponds to
conventional kinetic term, and the Lagrangian describes a theory of massless scalar field in
this case. Thus, $f_0(t)$ and $f_2(t)$ vanish at late times
\begin{equation*}
t \rightarrow +\infty:\quad f_0(t) = f_2(t) = 0,
\end{equation*}
We define $f_1(t)$ as a sum of two functions
\begin{equation}
\label{eq:f1_bounce}
f_1(t)=f_{10}(t) + f_{11}(t)
\end{equation}
with the following late-time asymptotics:
\begin{equation}
t \rightarrow +\infty:\quad f_{10}(t)=\dfrac{p}{t^2}, \quad f_{11}(t)=O(t^{-3}),
\label{may2-17-100}
\end{equation}
where $p$ is a constant coefficient. Let us consider asymptotic behavior of Einstein equations~\eqref{eq:einst_eq_setup} at $t \rightarrow +\infty$. The only non-vanishing terms are $H(t)$, $\dot{H}(t)$ and $f_{10}(t)$. By substituting
$H(t)=1/3t$ and $f_{10}(t)=p/t^2$ into Eq.~\eqref{eq:einst_eq_setup} with all other Lagrangian functions equal to zero
one finds that $p = 1/3$, which is a standard result in the gauge $\pi = t$. So, we choose $f_{10}(t)$ 
equal to
\begin{equation}
\label{eq:f10_bounce}
f_{10}(t) = \dfrac{1}{3(1 + t^2)}
\end{equation}
at all times. Note that $f_{11} (t)$ remains undefined, modulo its rapid decay as $t\rightarrow +\infty$, see~\eqref{may2-17-100}.
	
Let us now consider Einstein equations at all times. Substituting eqs.~\eqref{eq:gal_functions_bounce}
and~\eqref{eq:f1_bounce} into~\eqref{eq:einst_eq_setup} and solving for $f_{11}(t)$ and $f_2(t)$, we find
\begin{subequations}
\label{eq:f11_f2_bounce}
\begin{align}
f_{11}(t) =
&-2f_0 - f_{10} + \dot{k}_1 - 3\ddot{g}_{40} - 3\ddot{g}_{41} + 3H\left(k_1 - 3\dot{g}_{40} - \dot{g}_{41} - 2\dot{f}_{40}\right)\\
&\nonumber-6H^2\left(1 +2g_{40} - 3g_{41} + 4f_{40}\right) - 3\dot{H}\left(1 + 2g_{40} - 2g_{41} + 2f_{40}\right),\\
f_2(t)=
&\nonumber f_0  + \ddot{g}_{40} + \ddot{g}_{41} - H\left(3k_1 - 5 \dot{g}_{40} - 7\dot{g}_{41} - 2\dot{f}_{40}\right)\\
&+ 3H^2\left(1 + 2g_{40} - 4g_{41} + 6f_{40}\right)+ \dot{H}\left(1 + 2g_{40} - 2g_{41} + 2f_{40}\right).
\end{align}
\end{subequations}
Now, it is not immediate to choose the remaining undefined function $f_0(t)$ to satisfy the 
conditions of ghost absence and subluminality. As an intermediate step, we require that
 $f_0(t)$ satisfies the equation
\begin{equation}
\label{eq:sigma_theta_bounce}
\Sigma = q(1-\tanh(t))\Theta^2,
\end{equation}
where $\Sigma$ is determined by~\eqref{eq:Sigma_coeff_setup} and $q$ is a constant coefficient. From~\eqref{eq:Scal_setup} and~\eqref{eq:sigma_theta_bounce}
one has
\begin{equation}
\label{eq:sigma_theta2_bounce}
\mathcal{G_S} = q(1-\tanh(t))\mathcal{\hat{G}_T}^2 + 3\mathcal{\hat{G}_T}.
\end{equation}
From the latter equation it follows that one can vary $\mathcal{G_S}$ by changing coefficient $q$ and,
therefore, by changing $f_0(t)$. We will make use of latter equation later. Solving Eq.~\eqref{eq:sigma_theta_bounce} for
$f_0(t)$, one has
\begin{equation}
\label{eq:f0_bounce}
\begin{aligned}
&f_0(t) = \dfrac{1}{4}\Big(\dot{k}_1 - 3\ddot{g}_{40} - 3\ddot{g}_{41} + 3H\left(k_1 - 5\dot{g}_{40} - \dot{g}_{41} - 2\dot{f}_{40}\right)\\
&- 3H^2\left(3 + 6g_{40} - 6g_{41} - 2f_{40}\right) - 3\dot{H}\left(1 + 2f_{40} + 2g_{40} - 2g_{41}\right)\\
&+ q(1-\tanh(t))\left(- k_1 + \dot{g}_{40} + 3\dot{g}_{41} + H\left(1 + 2g_{40} - 6g_{41} + 10f_{40}\right)\right)^2\Big).
\end{aligned}
\end{equation}
By making use of eqs.~\eqref{eq:hubble_bounce},~\eqref{eq:k1_g40_g41_f40_bounce},~\eqref{eq:f10_bounce},~\eqref{eq:f11_f2_bounce} and~\eqref{eq:f0_bounce} one can check that $f_0(t)$, $f_{11}(t)$ and
$f_2(t)$ are smooth functions that rapidly vanish at late times. The Lagrangian will be fully
defined if we specify the values of coefficients $c_1$, $c_2$, $c_3$, $c_4$ and $q$. Note that we have not yet established the absence of instabilities.

Before we turn to the stability conditions of scalar and tensor perturbations, let us recall
that we want positive $\Theta$ at all times. With chosen asymptotics of Lagrangian functions, 
one has $\Theta|_{t=+\infty} = H$ at late times. The theory describes Einstein--Hilbert gravity and free
massless scalar field at late times, therefore, $G_4(\pi,X) = \frac{1}{2}$, and $\mathcal{G_T}|_{t=+\infty} = 1$. At early times,
we require
\begin{equation}
\label{eq:theta_early_bounce}
\Theta|_{t=-\infty} > 0.
\end{equation}
Now, recall that we want $\mathcal{G_T}$ to turn sign at some point; thus, we require
\begin{equation}
\label{eq:gt_early_bounce}
\mathcal{G_T}|_{t=-\infty} < 0.
\end{equation}
Let us now turn to the stability conditions~\eqref{eq:stability_cond_setup}.
\begin{enumerate}
\item Consider stability conditions for tensor perturbations. In Einstein-Hilbert gravity
at late times, one has
\begin{equation}
\label{eq:tensor_perturb_stability_late_bounce}
\mathcal{\hat{G}_T}|_{t=+\infty} = \mathcal{F_T}|_{t=+\infty} = 1 \; .
\end{equation}
We require $\mathcal{\hat{G}_T}$ and $\mathcal{F_T}$ to be smooth monotonic functions of $t$ and satisfy the following inequalities at early times
\begin{equation}
\label{eq:tensor_perturb_stability_early_bounce}
\mathcal{\hat{G}_T}|_{t=-\infty} > \mathcal{F_T}|_{t=-\infty} > 0.
\end{equation}
If eqs.~\eqref{eq:tensor_perturb_stability_late_bounce} and~\eqref{eq:tensor_perturb_stability_early_bounce} are valid, then the smooth monotonic functions $\mathcal{\hat{G}_T}$ and $\mathcal{F_T}$ are
positive at all times. In addition, we will choose them in such a way that inequality
$\mathcal{\hat{G}_T} > \mathcal{F_T}$ is satisfied at all times as well.

\item We focus now on the stability conditions for scalar perturbations. One can check that 
inequality $\mathcal{F_S}|_{t=+\infty}>0$ is equivalent to $\dot{H}|_{t=+\infty}<0$ and is satisfied with chosen Hubble
parameter~\eqref{eq:hubble_bounce}. At early times, we require
\begin{equation}
\label{eq:fs_early_bounce}
\mathcal{F_S}|_{t=-\infty} > 0.
\end{equation}

Let us leave aside for a moment the conditions of ghost absence $\mathcal{G_S} >0$ and
subluminality $\mathcal{G_S} > \mathcal{F_S}$. One possible choice of coefficients $c_1$, $c_2$, $c_3$, $c_4$, such that the  inequalities~\eqref{eq:theta_early_bounce},~\eqref{eq:gt_early_bounce},~\eqref{eq:tensor_perturb_stability_early_bounce} and~\eqref{eq:fs_early_bounce} are valid, is
\begin{equation}
\label{eq:c_coef_bounce}
c_1 = 1,\quad c_2 = -\dfrac{1}{4},\quad c_3 = \dfrac{1}{32},\quad c_4 = \dfrac{1}{4}.
\end{equation}
Now we recall Eq.~\eqref{eq:sigma_theta2_bounce} from which it follows that the condition of ghost absence $\mathcal{G_S} > 0$
is automatically satisfied when $\mathcal{\hat{G}_T} > 0$. Moreover, by taking
\begin{equation}
\label{eq:q_coef_bounce}
q = 20,
\end{equation}
the condition $\mathcal{G_S} > \mathcal{F_S}$ is satisfied as well.
\end{enumerate}

\begin{figure}[H] 
\vspace{-4ex} \centering
\subfigure[]{\includegraphics[width=0.5\linewidth]{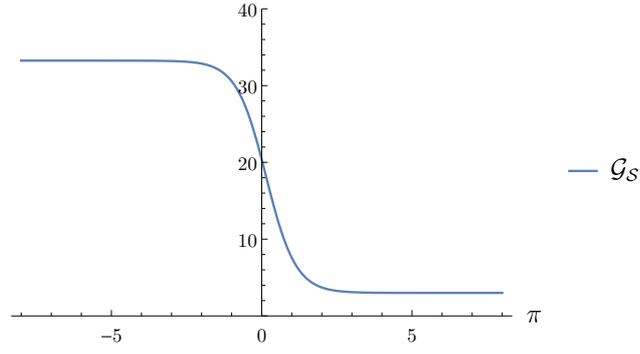} \label{pic:kinetic_scalar_bounce}}  
\hspace{4ex}
\subfigure[]{\includegraphics[width=0.5\linewidth]{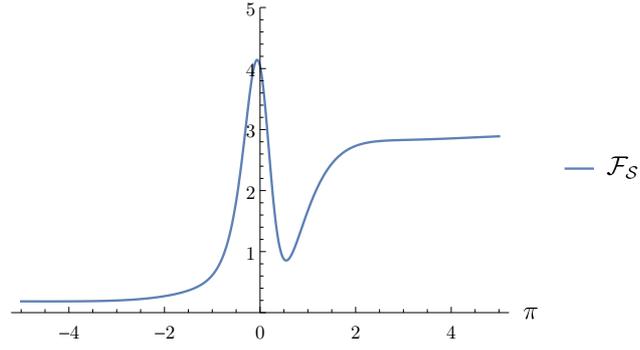} \label{pic:gradient_scalar_bounce}}
\hspace{4ex}
\subfigure[]{\includegraphics[width=0.5\linewidth]{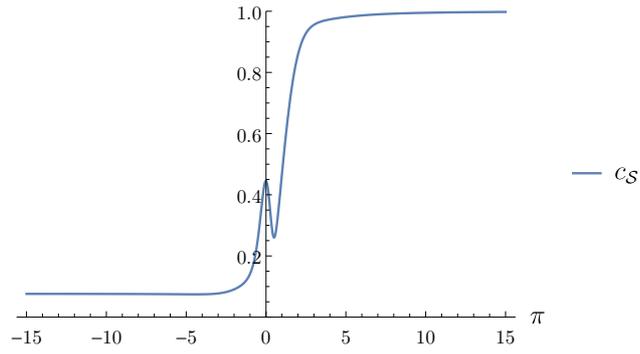} \label{pic:speed_scalar_bounce}}
\caption{Kinetic and gradient terms $\mathcal{G_S}$ (a) and $\mathcal{F_S}$ (b) and speed of scalar perturbations $c_\mathcal{S}$ (c) with parameters given by~\eqref{eq:c_coef_bounce},~\eqref{eq:q_coef_bounce}. 
Despite appearance,   $\mathcal{F_S}$ and  $c_\mathcal{S}$ are finite as $t\rightarrow -\infty$:  $\mathcal{F_S}  \approx 0.193$ at $t \rightarrow -\infty$;
$c_{\mathcal{S}} \approx 0.08$ at $t \rightarrow -\infty$.}
\label{pic:scalar_perturbations_bounce}
\end{figure}

\begin{figure}[H]
\vspace{-4ex} \centering
\subfigure[]{\includegraphics[width=0.5\linewidth]{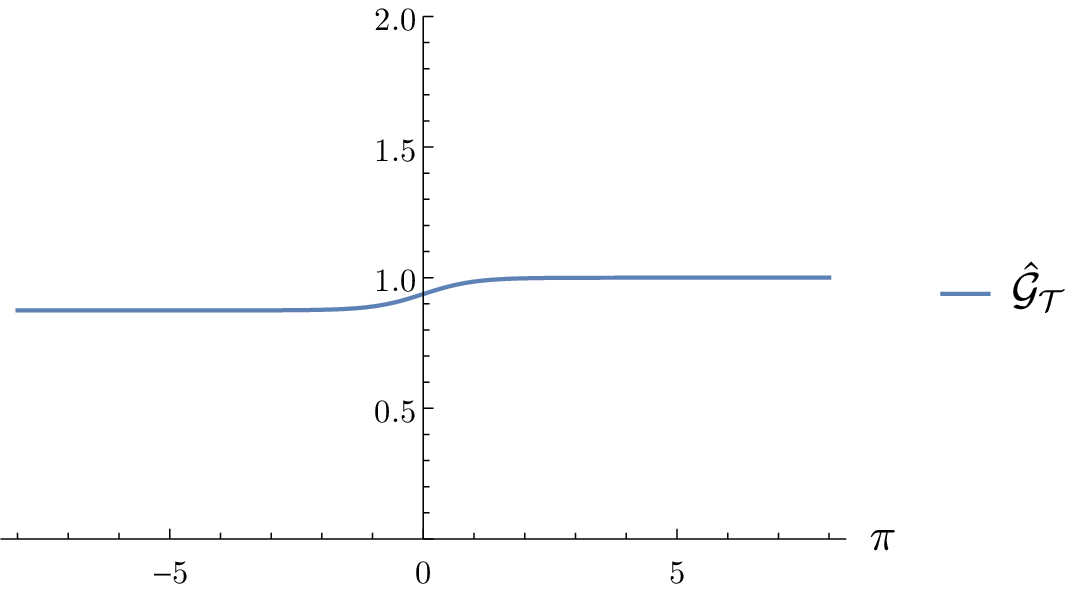} \label{pic:kinetic_tensor_bounce}}  \hspace{4ex}
\subfigure[]{\includegraphics[width=0.5\linewidth]{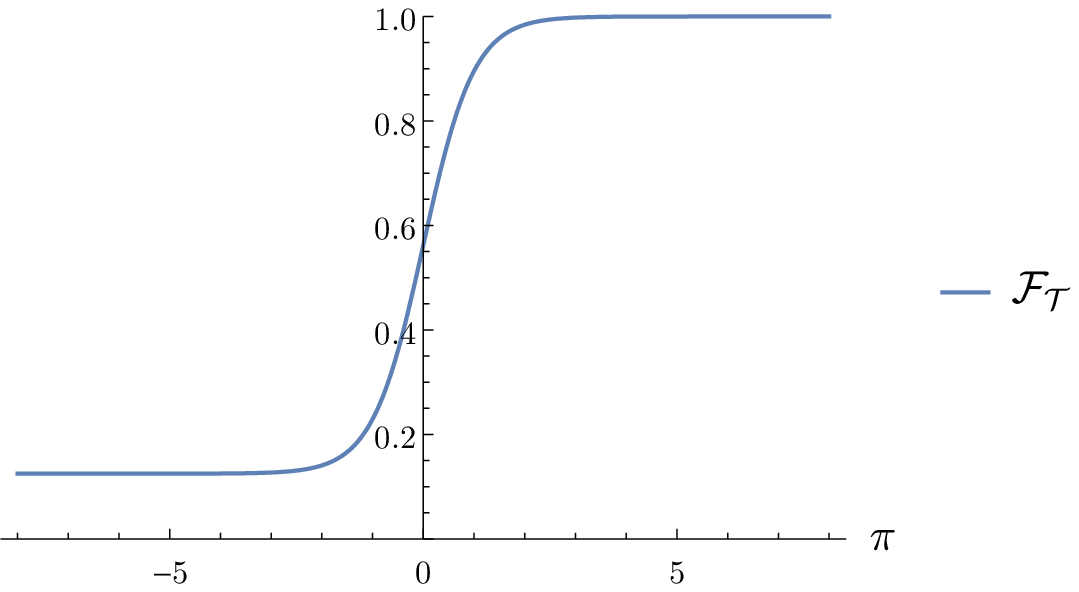} \label{pic:gradient_tensor_bounce}}
\hspace{4ex}
\subfigure[]{\includegraphics[width=0.5\linewidth]{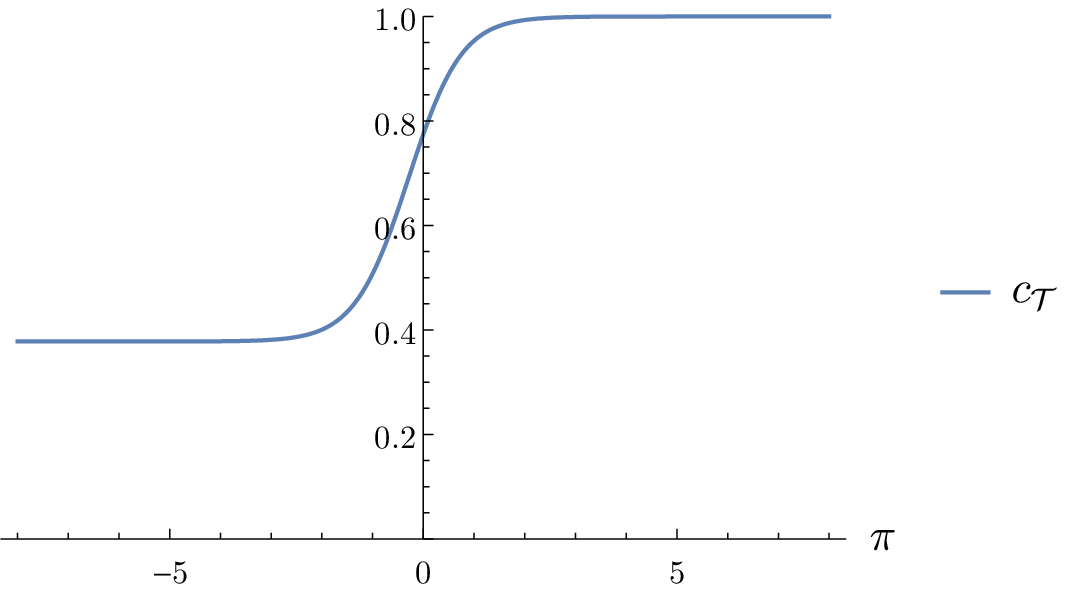} \label{pic:speed_tensor_bounce}}  
\caption{Kinetic and gradient terms for tensor perturbations, $\mathcal{\hat{G}_T}$ (a) and $\mathcal{F_T}$ (b), speed of
tensor perturbations $c_\mathcal{T}$ (c).}
\label{pic:tensor_perturbations_bounce} 
\end{figure}

\begin{figure}[H]
\center{\includegraphics[width=0.7\linewidth]{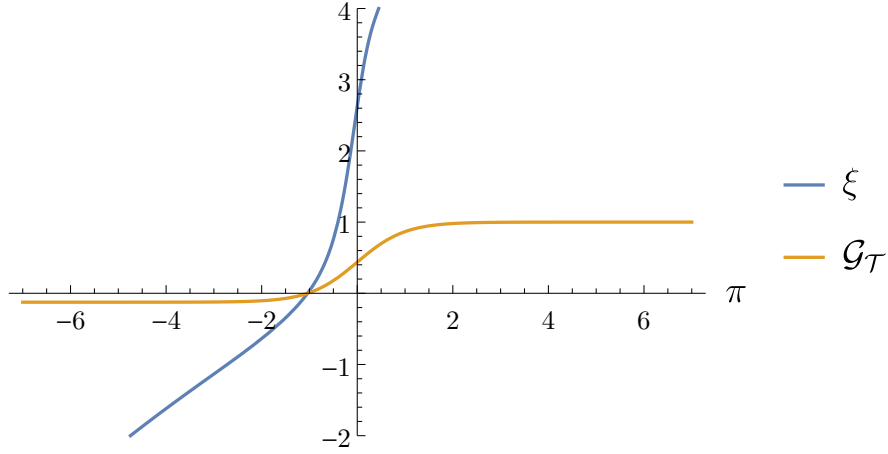} }
\caption{Evolution of $\xi$ and $\mathcal{G_T}$. $\xi$ = $\mathcal{G_T} = 0$ at $\pi = t = \artanh\left[\frac{7}{9}\right] \approx -1$.}
\label{pic:no_go}
\end{figure}

Thus, with parameters~\eqref{eq:c_coef_bounce},~\eqref{eq:q_coef_bounce} we have a completely healthy bounce in our beyond
Horndeski theory. Importantly, inequalities~\eqref{eq:theta_early_bounce},~\eqref{eq:gt_early_bounce},~\eqref{eq:tensor_perturb_stability_early_bounce} and~\eqref{eq:fs_early_bounce} with sufficiently large
$q$ guarantee the absence of pathologies only at early and late times. However, with the
coefficients $c_1$, $c_2$, $c_3$, $c_4$ and $q$ given by~\eqref{eq:c_coef_bounce} and~\eqref{eq:q_coef_bounce}, dynamics is stable throughout the
entire evolution. The behavior of scalar and tensor kinetic and gradient terms is shown in
Figs.~\ref{pic:scalar_perturbations_bounce} and~\ref{pic:tensor_perturbations_bounce}.
Both $\xi$ and $\mathcal{G_T}$ change sign simultaneously at $\pi = t = \artanh\left(\frac{7}{9}\right) \approx-1$, but $\mathcal{\hat{G}_T}$ stays
positive at all times, and our mechanism of evading the no-go theorem works. The behavior
of  $\xi$ and $\mathcal{G_T}$ is shown in Fig.~\ref{pic:no_go}.

Now let us turn to the Lagrangian functions. From eqs.~\eqref{eq:hubble_bounce},~\eqref{eq:gal_functions_bounce},~\eqref{eq:pi_bounce},~\eqref{eq:k1_g40_g41_f40_bounce},~\eqref{eq:f1_bounce},~\eqref{eq:f10_bounce},~\eqref{eq:f11_f2_bounce} and~\eqref{eq:f0_bounce} the Lagranginan functions have the following asymptotics at early times (we
no longer use the gauge $\dot{\pi}=1$):
\begin{subequations}
\begin{align*}
t\rightarrow -\infty~:~~
&F(\pi,X) = \mathcal{C}_0\cdot\dfrac{1}{\pi^2} + \left(\dfrac{1}{3} + \mathcal{C}_1\right)\dfrac{(\partial\pi)^2}{\pi^2}  + \mathcal{C}_2\dfrac{(\partial\pi)^4}{\pi^2},\\
&K(\pi,X) = 2c_1\dfrac{(\partial\pi)^2}{\pi},\\
&G_{4}(\pi,X) = \dfrac{1}{2} + 2c_2 + 2c_3(\partial\pi)^2,\\
&F_{4}(\pi,X) = 2c_4,
\end{align*}	
\end{subequations}
where the coefficients $\mathcal{C}_0$, $\mathcal{C}_1$ and $\mathcal{C}_2$ are
\begin{subequations}
\begin{align*}
\mathcal{C}_0 = &\dfrac{q}{18} \left(1 - 6 c_1 + 4 c_2 - 12 c_3 + 20 c_4\right)^2 + \dfrac{4}{3} c_4\\
\mathcal{C}_1 = &\dfrac{4}{3}(c_2 - c_4) - 2\mathcal{C}_0\\
\mathcal{C}_2 = &- \dfrac{2}{3} (3 c_1 + 2 c_3 - 4 c_4) + \mathcal{C}_0 \; .
\end{align*}
\end{subequations}
These can be found from eqs.~\eqref{eq:f11_f2_bounce} and~\eqref{eq:f0_bounce} at $t \rightarrow -\infty$. The Lagrangian at early times has
the form
\begin{equation*}
\begin{aligned}
\mathcal{L}|_{t=-\infty}
&=\mathcal{C}_0\cdot\dfrac{1}{\pi^2} + \left(\dfrac{1}{3} + \mathcal{C}_1\right)\dfrac{(\partial\pi)^2}{\pi^2}  + \mathcal{C}_2\dfrac{(\partial\pi)^4}{\pi^2}\\
&+2c_1\dfrac{(\partial\pi)^2}{\pi}\square\pi - \left(\dfrac{1}{2} + 2c_2 + 2c_3(\partial\pi)^2\right)R + 4c_3\left[(\square\pi)^2 - \nabla^{\mu\nu}\pi\nabla_{\mu\nu}\pi\right]\\
&+ 2c_4\epsilon^{\mu\nu\rho\sigma}{\epsilon^{\mu'\nu'\rho'}}_\sigma\nabla_\mu\pi\nabla_\mu'\pi\nabla_{\nu\nu'}\pi\nabla_{\rho\rho'}\pi.
\end{aligned}
\end{equation*}
Note that it contains both Horndeski and beyond Horndeski terms. On the other hand, at
late times the only non-vanishing function is $f_1(t)$. From eqs.~\eqref{eq:pi_bounce}, and~\eqref{eq:f10_bounce} at late times
one has
\begin{subequations}
\begin{align*}
t\rightarrow +\infty~:~~
&F(\pi,X) = \dfrac{1}{3}\dfrac{(\partial\pi)^2}{\pi^2},\\
&K(\pi,X) = 0,\\
&G_{4}(\pi,X) = \dfrac{1}{2},\\
&F_{4}(\pi,X) = 0.
\end{align*}	
\end{subequations}
The Lagrangian at $t=+\infty$ has the form
\begin{equation*}
\mathcal{L}|_{t=+\infty}=-\dfrac{1}{2}R + \dfrac{1}{3}\dfrac{(\partial\pi)^2}{\pi^2}=-\dfrac{1}{2}R + \dfrac{1}{3}(\partial\phi)^2,
\end{equation*}
where $\phi = \ln(\pi)$. As we anticipated, Galileon field becomes a free massless scalar field 
interacting with the Einstein-Hilbert gravity. 

\begin{figure}[H]
\begin{minipage}[h]{0.4\linewidth}
\center{\includegraphics[width=1\linewidth]{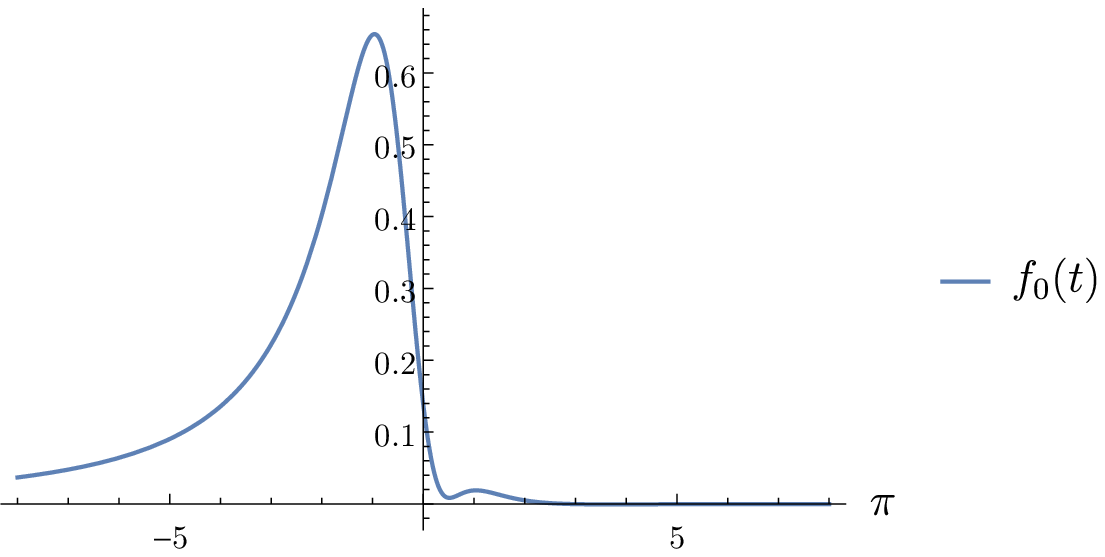}} (a) \\
\end{minipage}
\hfill
\begin{minipage}[H]{0.4\linewidth}
\center{\includegraphics[width=1\linewidth]{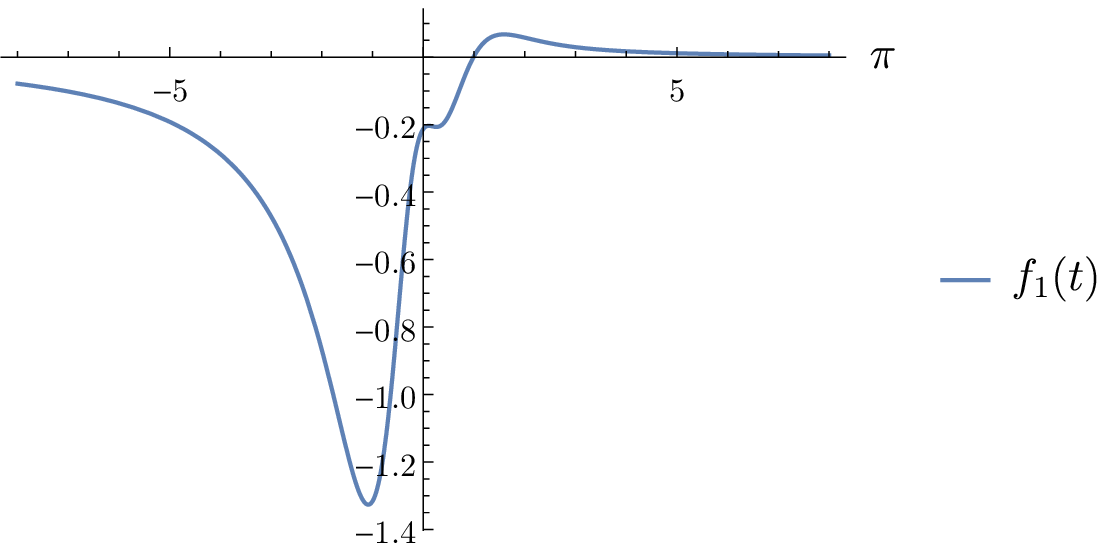}} (b) \\
\end{minipage}
\vfill
\begin{minipage}[H]{0.4\linewidth}
\center{\includegraphics[width=1\linewidth]{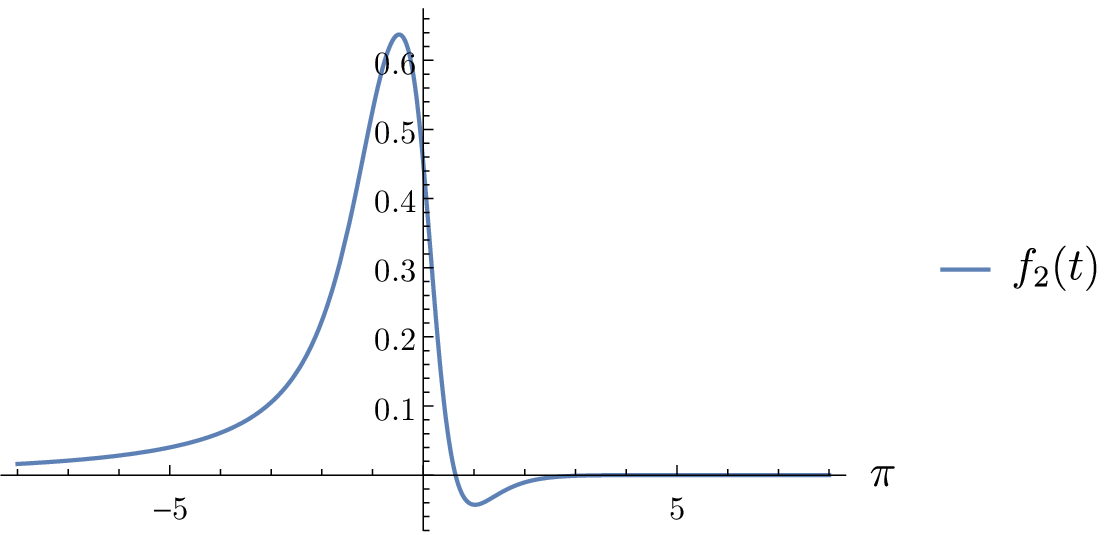}} (c) \\
\end{minipage}
\hfill
\begin{minipage}[H]{0.4\linewidth}
\center{\includegraphics[width=1\linewidth]{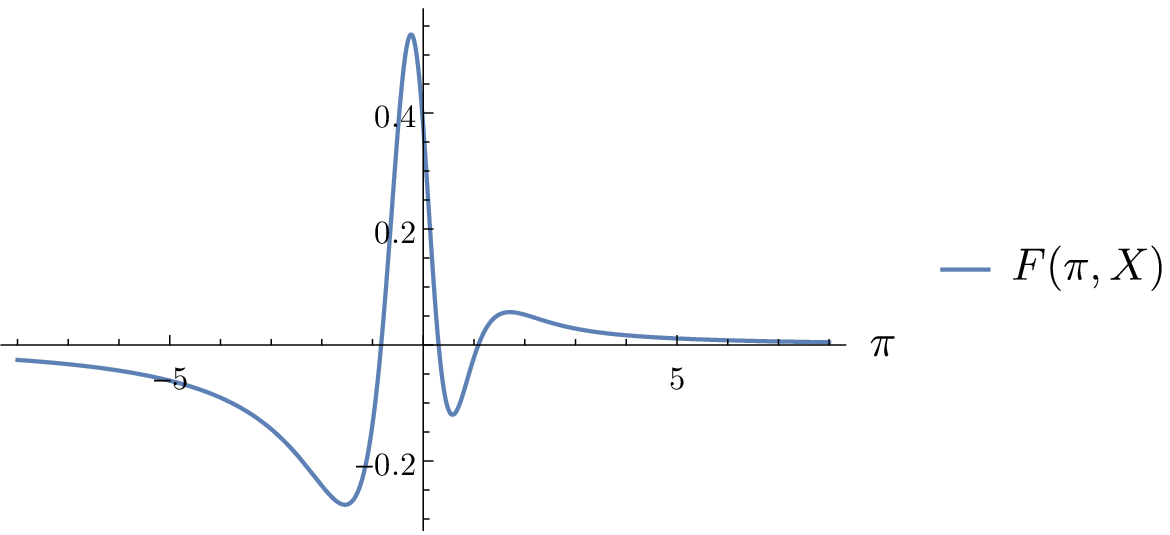}} (d) \\
\end{minipage}
\centering
\caption{Lagrangian functions $f_0(\pi)$, 
$f_1(\pi)$, $f_2(\pi)$ and $F(\pi,X)$ from~\eqref{eq:gal_functions_bounce} in the gauge $\dot{\pi}=1$. Note that all functions are smooth and without singularities at any 
point during the entire evolution.}
\label{pic:gal_functions}
\end{figure}

To discuss the overall behavior of the Lagrangian functions, we recall that $k_1 (\pi)$, $g_{40} (\pi)$,
$g_{41} (\pi)$, and $f_{40} (\pi)$ are explicitly given by~\eqref{eq:k1_g40_g41_f40_bounce}, with $t$ substituted by $\pi$ (we are back to the
gauge $\dot{\pi}=1$). The rest of the functions (namely, $f_0$, $f_1$, $f_2$) are shown in Fig.~\ref{pic:gal_functions}, where
constant coefficients are chosen as in eqs.~\eqref{eq:c_coef_bounce} and~\eqref{eq:q_coef_bounce}.
	
The theory possesses one peculiar feature: it is scale covariant at $t=-\infty$. This scale 
covariance is behind the simple asymptotic power-law behavior at early times. Let us make 
a scale transformation of the coordinates
\begin{equation}
\label{eq:scale_transform_bounce}
x^\mu \rightarrow \lambda x^\mu.
\end{equation}
The Galileon field $\pi(x^\mu)$ and metric $g_{\mu\nu}(x^\mu)$  transform as follows:
\begin{subequations}
\label{eq:solution_scale_transfrom_bounce}
\begin{align}
&\pi'(x^\mu) = \lambda^{-\frac{3}{2}}\pi(\lambda x^\mu),\\
&g'_{\mu\nu}(x^\mu) = \lambda^{-1}g_{\mu\nu}(\lambda x^\mu),
\end{align}
\end{subequations}
and, respectively, $g'^{\mu\nu}(x^\mu) = \lambda g^{\mu\nu}(\lambda x^\mu)$. One can check that the early time Lagrangian
transforms in the following way:
\begin{equation*}
\mathcal{L}'(x^\mu) = \lambda^{3}\mathcal{L}(\lambda x^\mu),
\end{equation*}
and the equations of motion remain unchanged. Moreover, the solution is scale invariant in
terms of conformal time $\eta = \int \frac{\mathrm{d}t}{a(t)}$. Indeed, as $t \rightarrow -\infty$, the asymptotics of the solution is
\begin{equation}
\label{eq:solution_conftime_bounce}
t \rightarrow -\infty:\quad \pi \sim -|\eta|^{\frac{2}{3}}, \quad g_{\mu\nu} \sim|\eta|\cdot\eta_{\mu\nu},
\end{equation}
where $\eta_{\mu\nu}$ is the Minkowski metric. From eqs.~\eqref{eq:solution_scale_transfrom_bounce} and~\eqref{eq:solution_conftime_bounce} it follows that the solution is
invariant under the transformations~\eqref{eq:scale_transform_bounce} at early times. 

\section{Genesis: an example}
\label{sec:genesis}
We now apply the same technique to construct stable, geodesically complete Genesis solution.
The only thing that we have to change as compared to the previous reasoning is the Hubble
parameter and function $k_1(t)$, since the latter is proportional to the Hubble parameter at
early times.

\begin{figure}[H]
\center{\includegraphics[width=0.7\linewidth]{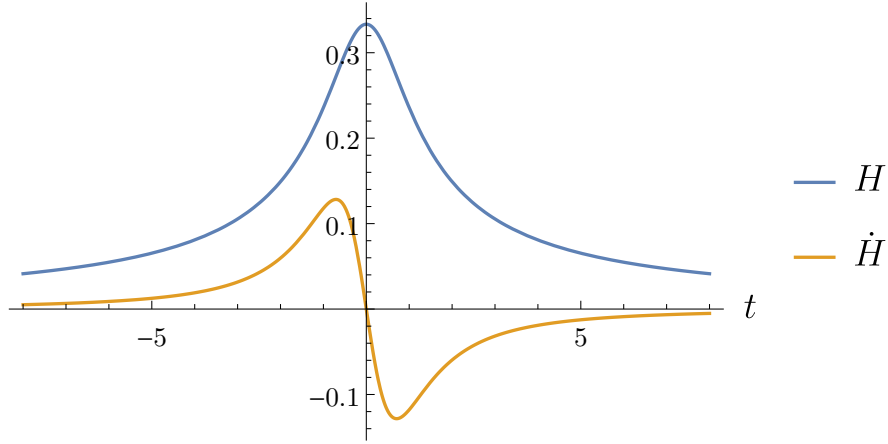} }
\caption{Hubble parameter $H$ and its time derivative $\dot{H}$.}
\label{pic:hubble_genesis}
\end{figure}

Like in the case of the bounce, we require the Galileon to become a conventional massless
scalar field at late times. Thus, the late-time asymptotic of the Hubble parameter is
\begin{equation*}
t \rightarrow +\infty:\quad H(t) = \dfrac{1}{3t}.
\end{equation*}
We choose the Hubble parameter at all times equal to
\begin{equation*}
H(t) = \dfrac{1}{3\sqrt{1 + t^2}}.
\end{equation*}
For the scale factor $a(t)$ one has
\begin{equation*}
a(t) = \left[t +\sqrt{1 + t^2}\right]^\frac{1}{3}.
\end{equation*}
The evolution of the Hubble parameter and its time derivative are shown in Fig.~\ref{pic:hubble_genesis}.

\begin{figure}[H]
\begin{minipage}[h]{0.4\linewidth}
\center{\includegraphics[width=1\linewidth]{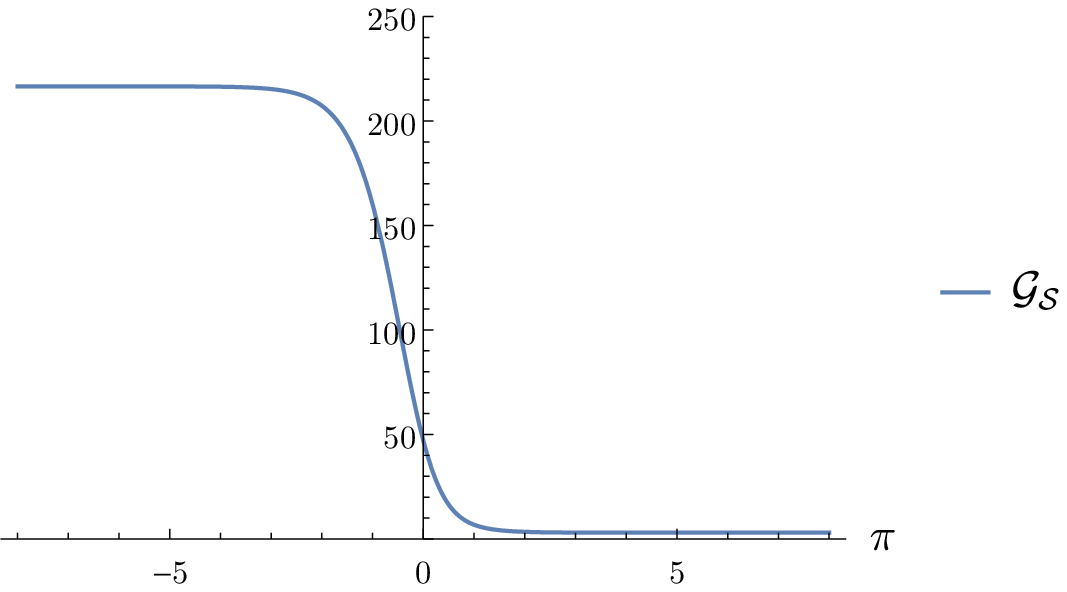}} (a) \\
\end{minipage}
\hfill
\begin{minipage}[H]{0.4\linewidth}
\center{\includegraphics[width=1\linewidth]{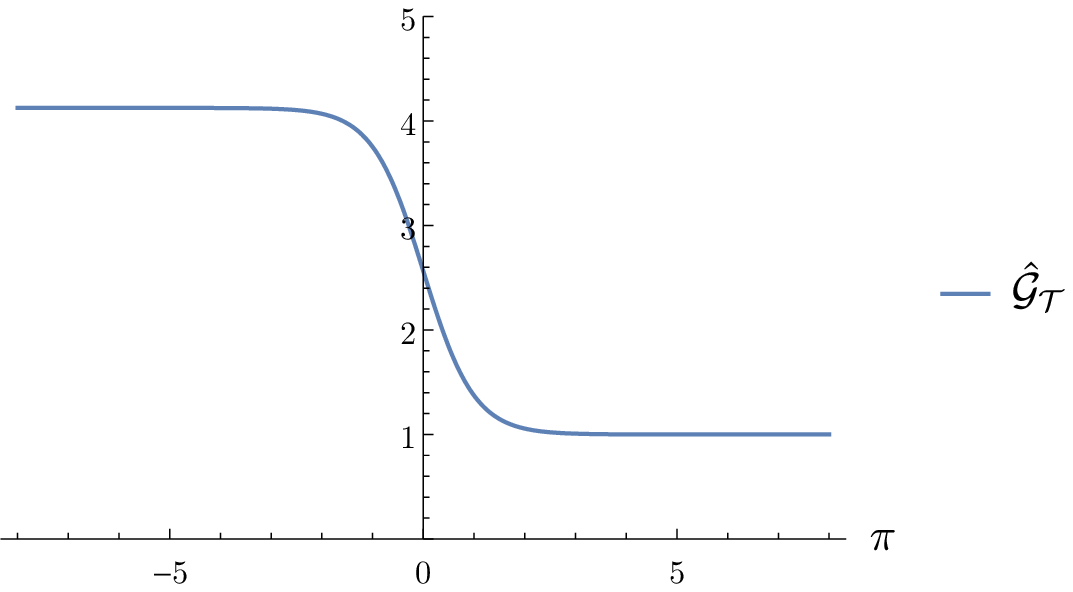}} (d) \\
\end{minipage}
\vfill
\begin{minipage}[H]{0.4\linewidth}
\center{\includegraphics[width=1\linewidth]{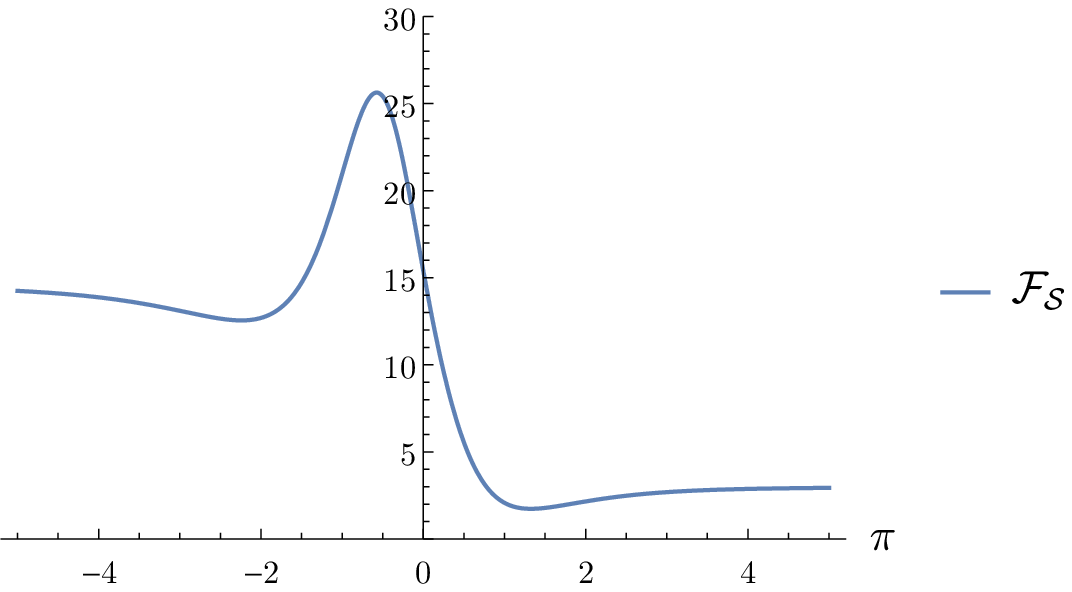}} (b) \\
\end{minipage}
\hfill
\begin{minipage}[H]{0.4\linewidth}
\center{\includegraphics[width=1\linewidth]{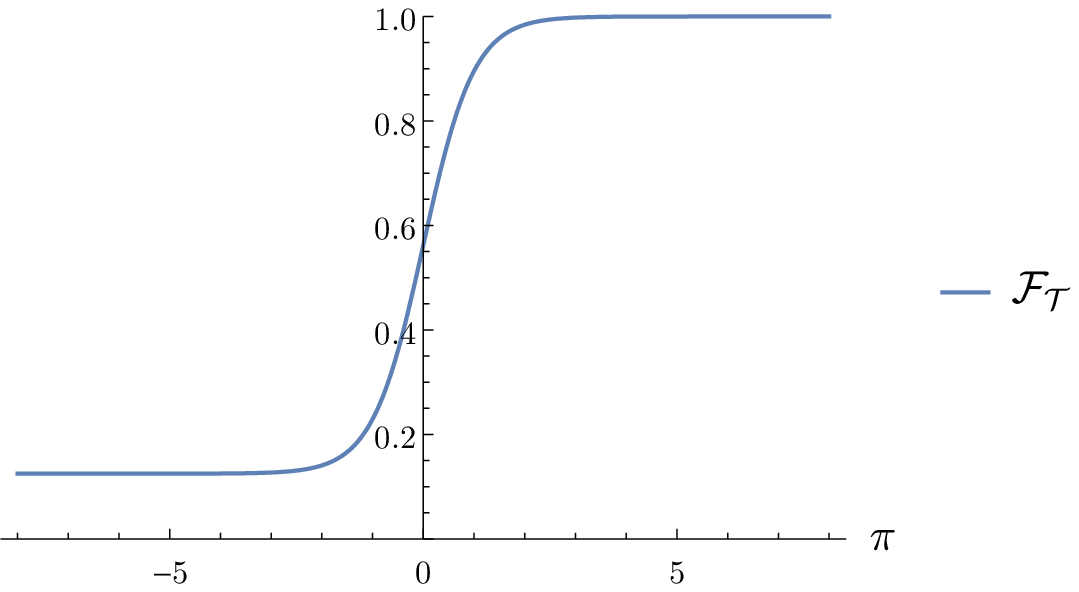}} (e) \\
\end{minipage}
\vfill
\begin{minipage}[H]{0.4\linewidth}
\center{\includegraphics[width=1\linewidth]{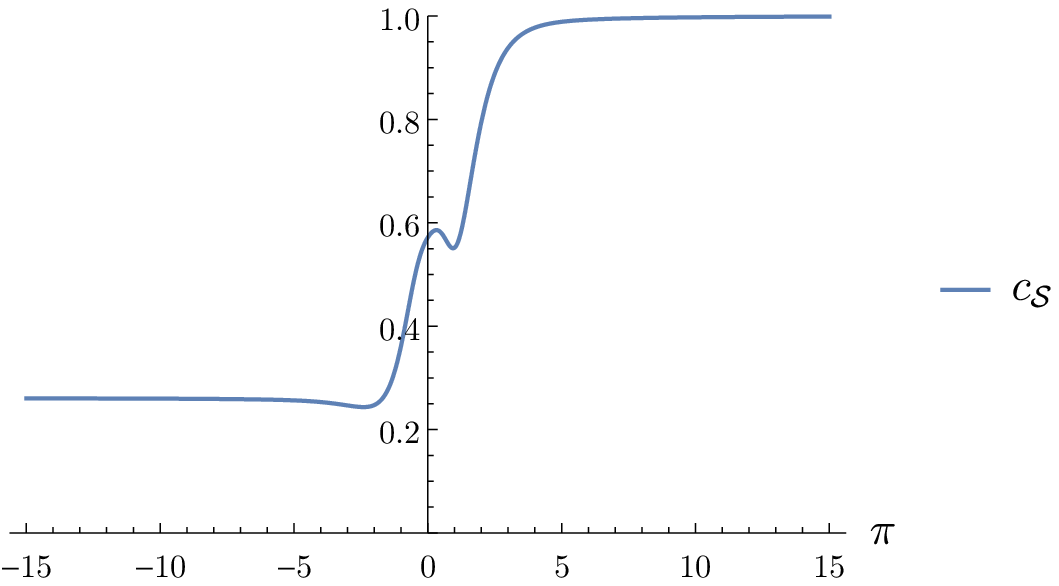}} (c) \\
\end{minipage}
\hfill
\begin{minipage}[H]{0.4\linewidth}
\center{\includegraphics[width=1\linewidth]{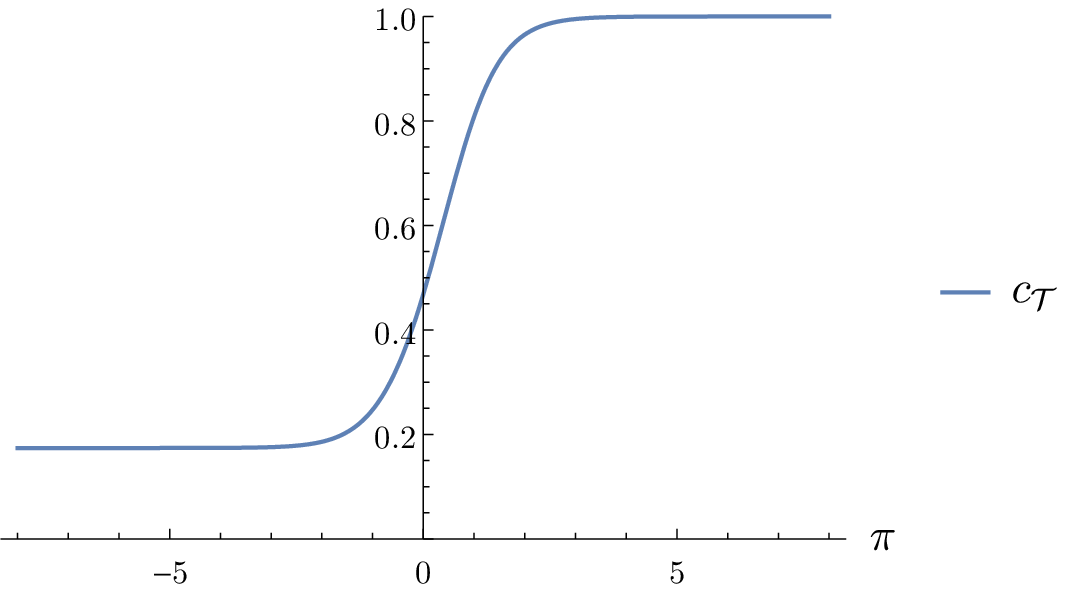}} (f) \\
\end{minipage}
\caption{Kinetic and gradient terms $\mathcal{G_S}$ (a) and $\mathcal{F_S}$ (b), speed of scalar perturbations $c_\mathcal{S}$ (c). Kinetic and gradient terms $\mathcal{\hat{G}_T}$ (d) and $\mathcal{F_T}$ (e), speed of tensor perturbations $c_\mathcal{T}$ (f).}
\label{pic:perturbations_genesis}
\end{figure}

As we have argued earlier, we only have to change $k_1(t)$. In the case of bounce it was
chosen to be proportional to the Hubble parameter at early times, and to vanish at late
times. We make a similar Ansatz in the Genesis case, and  choose $k_1(t)$ in the following form:
\begin{equation*}
k_1(t) = c_1\dfrac{1}{\sqrt{1+t^2}}\left(1-\tanh (t)\right).
\end{equation*}
The Ansatz for the the functions $g_{4 0}$, $g_{4 1}$ and $f_{4 0}$ is completely the same as in~\eqref{jun2-17-2},~\eqref{jun2-17-3}, and we again request that Eq.~\eqref{eq:sigma_theta_bounce} is satisfied. Then the functions $f_{11}$, $f_2$ and $f_0$ are again
found from~\eqref{eq:f11_f2_bounce},~\eqref{eq:f0_bounce}.
It remains to make a choice of constants; our example is
\begin{equation}
\label{eq:c_coef_genesis}
c_1 = \dfrac{31}{8},\quad c_2 = -\dfrac{3}{8},\quad c_3 = \dfrac{5}{32},\quad c_4 = \dfrac{21}{16}
\end{equation}
and
\begin{equation}
\label{eq:q_coef_genesis}
q = 6.
\end{equation}
With this choice, the Genesis solution is completely healthy. The evolution of kinetic and
gradient terms as well as the speed of scalar and tensor perturbations are presented in Fig.~\ref{pic:perturbations_genesis}.

\begin{figure}[H]
\center{\includegraphics[width=0.7\linewidth]{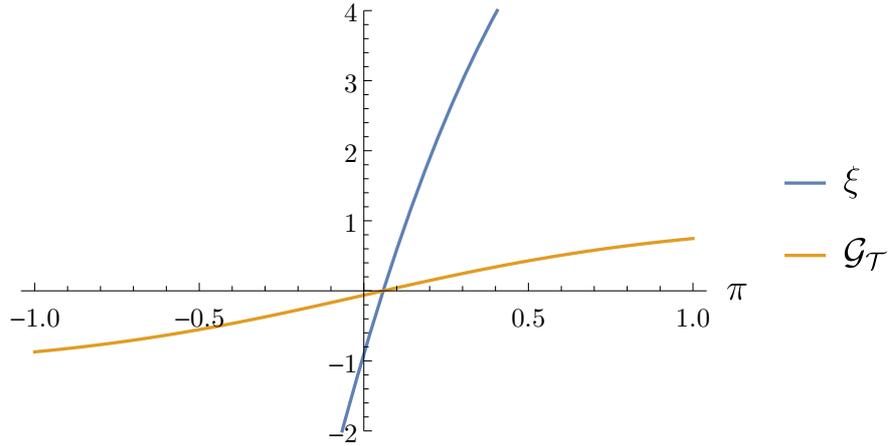} }
\caption{Evolution of $\xi$ and $\mathcal{G_T}$. Zero crossing
 $\xi$ = $\mathcal{G_T} = 0$ occurs at $\pi = t = \artanh\left[\frac{1}{17}\right] \approx 0.06$.}
\label{pic:no_go_genesis}
\end{figure}

\begin{figure}[H]
\begin{minipage}[h]{0.4\linewidth}
\center{\includegraphics[width=1\linewidth]{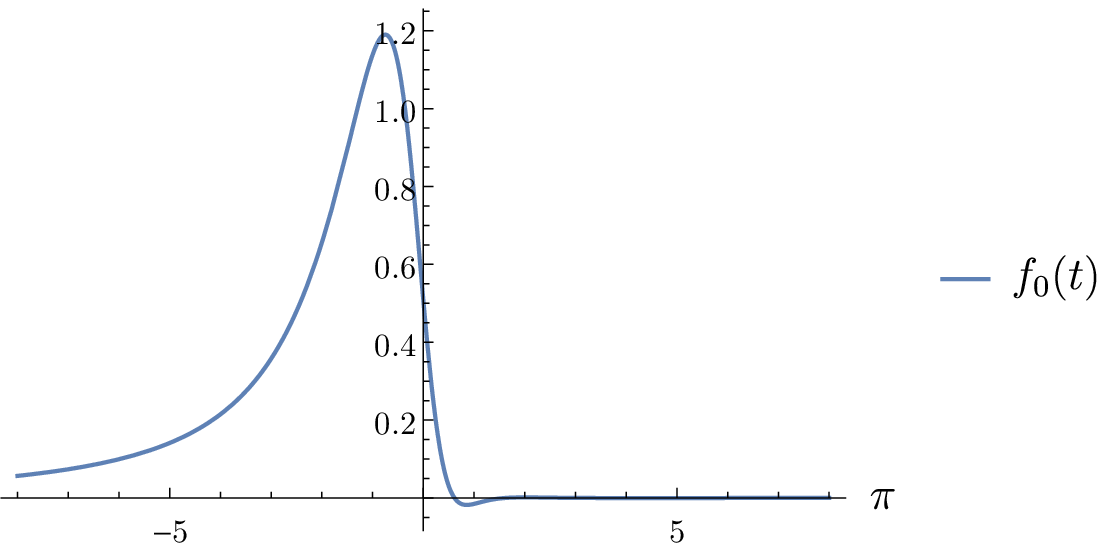}} (a) \\
\end{minipage}
\hfill
\begin{minipage}[H]{0.4\linewidth}
\center{\includegraphics[width=1\linewidth]{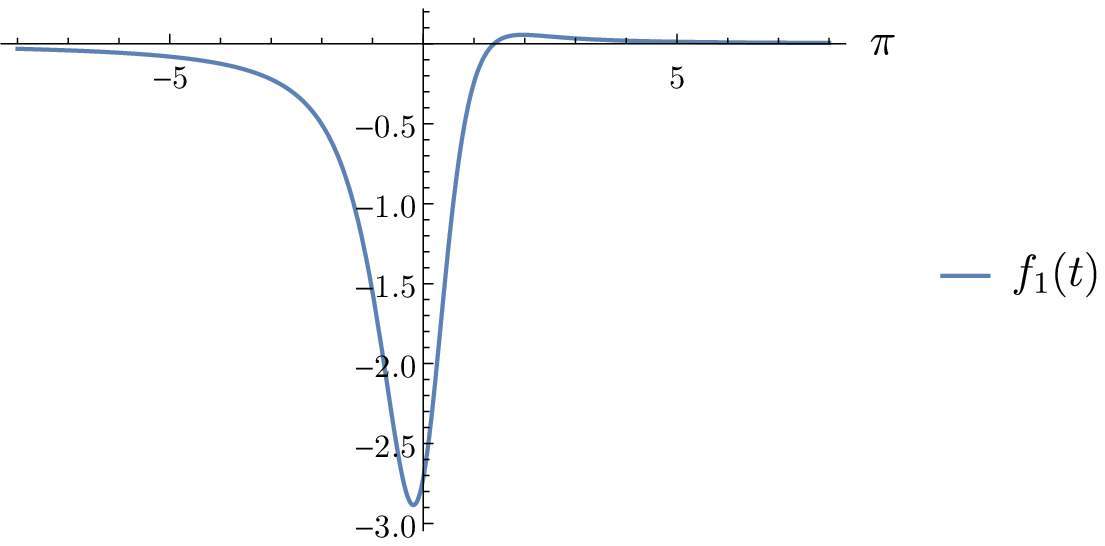}} (b) \\
\end{minipage}
\vfill
\begin{minipage}[H]{0.4\linewidth}
\center{\includegraphics[width=1\linewidth]{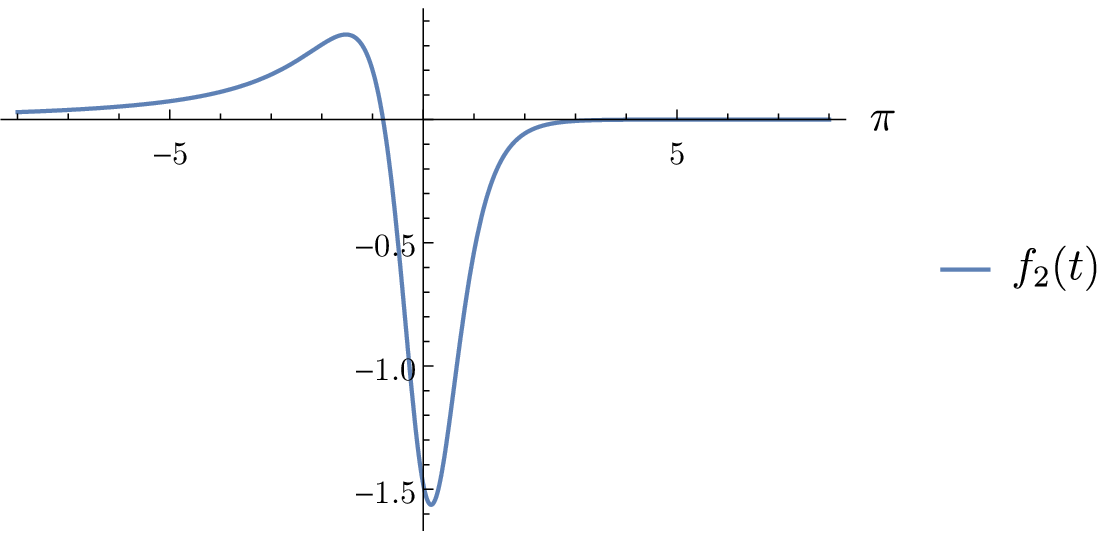}} (c) \\
\end{minipage}
\hfill
\begin{minipage}[H]{0.4\linewidth}
\center{\includegraphics[width=1\linewidth]{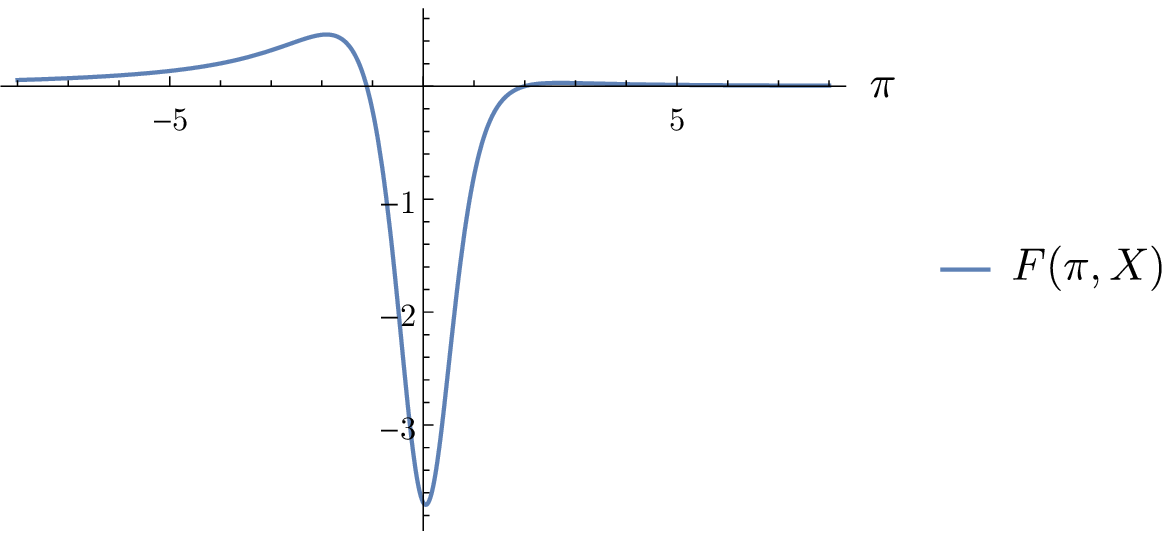}} (d) \\
\end{minipage}
\centering
\caption{Lagrangian functions $f_0(\pi)$, $f_1(\pi)$, $f_2(\pi)$ and $F(\pi,X)$ in the gauge $\dot{\pi}=1$. Note that all functions are smooth and without singularities at any point during the entire evolution.}
\label{pic:gal_functions_genesis}
\end{figure}

To complete our discussion of the Genesis solution, we show in  Fig.~\ref{pic:no_go_genesis} the behavior of  $\xi$
and $\mathcal{G_T}$, which enables one to evade the no-go argument, cf. Fig.~\ref{pic:no_go}. The functions $\xi (t)$ and
$\mathcal{G_T}$ cross zero at $\pi = t = \artanh\left[\frac{1}{17}\right] \approx 0.06$. The functions $f_0$, $f_1$, $f_2$ are shown in Fig.~\ref{pic:gal_functions_genesis},
where the constant coefficients are chosen as in eqs.~\eqref{eq:c_coef_genesis} and~\eqref{eq:q_coef_genesis}. We see that the Genesis
solution is completely healthy, with all Lagranian functions smooth at all values of $\pi$.

Finally, let us consider asymptotic completeness of our Genesis solution. Since the
asymptotics of the Lagrangian functions are similar to those in Sec.~\ref{sec:bounce}, our model of Genesis is also
scale covariant at $t\rightarrow -\infty$. Recall now that the scale factor for our scenario has the form
\begin{equation*}
a(t)=\left[t+\sqrt{1+t^2}\right]^{\frac{1}{3}}.
\end{equation*}
At early times ($t \rightarrow -\infty$), the scale factor has the following asymptotics:
\begin{equation}
\label{eq:scale_geodesic}
a(t) = \dfrac{1}{\left(-2t\right)^\frac{1}{3}}.
\end{equation}
Let us consider the null and timelike geodesics
\begin{subequations}
\label{eq:null_geodesic}
\begin{align}
&t'^2 = a(t)^2x'^2,\\
&t'^2 = a(t)^2x'^2+1,
\end{align}
\end{subequations}
where prime denotes the derivative with respect to affine parameter $\tau$. We keep only one
coordinate $x$, since the geodesic is a straight line. One of the two geodesic equations has the
form 
\begin{equation*}
\left(a(t)^2 x'\right)'=0,
\end{equation*}
which implies that
\begin{equation*}
a(t)^2 x' = const = C,
\end{equation*}
where we can set $C = 1$ without loss of generality. From the latter equation and eqs.~\eqref{eq:null_geodesic}
one obtains the following expressions for the null geodesic
\begin{subequations}
\begin{align}
\label{eq:parameter_geodesic_null}
&\tau = \int	a(t) \mathrm{d}t,\\
\label{eq:coordinate_geodesic_null}
&x = \int	\frac{1}{a(t)} \mathrm{d}t,
\end{align}
\end{subequations}
and the timelike one
\begin{subequations}
\begin{align}
\label{eq:parameter_geodesic_time}
&\tau = \int	\frac{a(t)}{\sqrt{1+a(t)^2}} \mathrm{d}t,\\
\label{eq:coordinate_geodesic_time}
&x = \int	\frac{\sqrt{1+a(t)^2}}{a(t)} \mathrm{d}t.
\end{align}
\end{subequations}
We see that~\eqref{eq:parameter_geodesic_time} and~\eqref{eq:coordinate_geodesic_time} have same asymptotic behavior as~\eqref{eq:parameter_geodesic_null} and~\eqref{eq:coordinate_geodesic_null}, since
$a(t)\rightarrow 0$ and, therefore, $\sqrt{1+a(t)^2}\rightarrow 1$ for $t\rightarrow -\infty$. Hence, we can study a simpler set of
equations~\eqref{eq:parameter_geodesic_null} and~\eqref{eq:coordinate_geodesic_null}. From eqs.~\eqref{eq:scale_geodesic},~\eqref{eq:parameter_geodesic_null} and~\eqref{eq:coordinate_geodesic_null} we get
\begin{subequations}
\begin{align*}
-\tau \sim (-t)^\frac{2}{3},\\
-x \sim (-t)^\frac{4}{3}.
\end{align*}
\end{subequations}
Moving to the past along the geodesic ($t\rightarrow -\infty$), the affine parameter tends to infinity
($\tau \rightarrow -\infty$), and so does the coordinate ($x\rightarrow -\infty$).  Hence, at no value of the affine
parameter we leave the FRWL coordinate patch. Thus, our Genesis model is geodesically
complete.

\section{Conclusion}
\label{sec:conclusion}
In this paper we constructed an explicit ``classical'' cosmological bounce that is free of any
kind of instabilities and singularities during the whole evolution. We also gave an  example of
fully stable and geodesically complete Genesis. We presented the Lagrangian functions and
checked that the Einstein  and field equations are satisfied. We also found that the simple
form of the solution at early times is the result of the asymptotic scale covariance of the
theory. The characteristic feature of the solutions, namely, the flow of the Galileon field 
into a conventional massless scalar field at late times, enables one to potentially merge the bouncing and/or Genesis scenario with the conventional evolution at later stages.

\section*{Acknowledgements}
We are indebted to V. Rubakov for fruitful discussions and thoughtful reading of the manuscript. 
This work has been supported by Russian Science Foundation grant 14-22-00161.


\begin{thebibliography}{99}
\bibitem{Horndeski:1974wa}
G.~W.~Horndeski,
``Second-order scalar-tensor field equations in a four-dimensional space,''
\href{http://dx.doi.org/10.1007/BF01807638}{Int.\ J.\ Theor.\ Phys.\  {\bf 10} (1974) 363}.

\bibitem{Fairlie:1991qe}
D.~B.~Fairlie, J.~Govaerts and A.~Morozov,
``Universal field equations with covariant solutions,''
\href{http://dx.doi.org/10.1016/0550-3213(92)90455-K}{Nucl.\ Phys.\ B {\bf 373} (1992) 214}
\href{https://arxiv.org/pdf/hep-th/9110022.pdf}{[hep-th/9110022]}.
		
D.~B.~Fairlie and J.~Govaerts,
``Universal field equations with reparametrization invariance,''
\href{http://dx.doi.org/10.1016/0370-2693(92)90273-7}{Phys.\ Lett.\ B {\bf 281} (1992) 49}.
\href{https://arxiv.org/pdf/hep-th/9202056.pdf}{[hep-th/9202056]}.

D.~B.~Fairlie and J.~Govaerts,
``Euler hierarchies and universal equations,''
\href{http://dx.doi.org/10.1063/1.529904}{J.\ Math.\ Phys.\  {\bf 33} (1992) 3543}
\href{https://arxiv.org/pdf/hep-th/9204074.pdf}{[hep-th/9204074]}.

\bibitem{Luty:2003vm}
M.~A.~Luty, M.~Porrati and R.~Rattazzi,
``Strong interactions and stability in the DGP model,''
\href{http://dx.doi.org/10.1088/1126-6708/2003/09/029}{JHEP {\bf 0309} (2003) 029}
\href{https://arxiv.org/pdf/hep-th/0303116.pdf}{[hep-th/0303116]}.

A.~Nicolis and R.~Rattazzi,
``Classical and quantum consistency of the DGP model,''
\href{http://dx.doi.org/10.1088/1126-6708/2004/06/059}{JHEP {\bf 0406} (2004) 059}
\href{https://arxiv.org/pdf/hep-th/0404159.pdf}{[hep-th/0404159]}.
	
\bibitem{Nicolis:2008in}
A.~Nicolis, R.~Rattazzi and E.~Trincherini,
``The Galileon as a local modification of gravity,''
\href{http://dx.doi.org/10.1103/PhysRevD.79.064036}{Phys.\ Rev.\ D {\bf 79} (2009) 064036}
\href{https://arxiv.org/pdf/0811.2197.pdf}{[arXiv:0811.2197 [hep-th]]}.
	
\bibitem{Deffayet:2010zh}
C.~Deffayet, S.~Deser and G.~Esposito-Farese,
``Arbitrary $p$-form Galileons,''
\href{http://dx.doi.org/10.1103/PhysRevD.82.061501}{Phys.\ Rev.\ D {\bf 82} (2010) 061501}
\href{https://arxiv.org/pdf/1007.5278.pdf}{[arXiv:1007.5278 [gr-qc]]}.

\bibitem{Deffayet:2010qz}
C.~Deffayet, O.~Pujolas, I.~Sawicki and A.~Vikman,
``Imperfect Dark Energy from Kinetic Gravity Braiding,''
\href{http://dx.doi.org/10.1088/1475-7516/2010/10/026}{JCAP {\bf 1010} (2010) 026}
\href{https://arxiv.org/pdf/1008.0048.pdf}{[arXiv:1008.0048 [hep-th]]}.

\bibitem{Kobayashi:2011nu}
T.~Kobayashi, M.~Yamaguchi and J.~Yokoyama,
``Generalized G-inflation: Inflation with the most general second-order field equations,''
\href{http://dx.doi.org/10.1143/PTP.126.511}{Prog.\ Theor.\ Phys.\  {\bf 126} (2011) 511}
\href{https://arxiv.org/pdf/1105.5723.pdf}{[arXiv:1105.5723 [hep-th]]}.

\bibitem{Padilla:2012dx}
A.~Padilla and V.~Sivanesan,
``Covariant multi-galileons and their generalisation,''
\href{http://dx.doi.org/10.1007/JHEP04(2013)032}{JHEP {\bf 1304} (2013) 032}
\href{https://arxiv.org/pdf/1210.4026.pdf}{[arXiv:1210.4026 [gr-qc]]}.

\bibitem{Rubakov:2014jja}
V.~A.~Rubakov,
``The Null Energy Condition and its violation,''
\href{http://dx.doi.org/10.3367/UFNe.0184.201402b.0137}{Phys.\ Usp.\  {\bf 57} (2014) 128
[Usp.\ Fiz.\ Nauk {\bf 184} (2014) no.2,  137]}
\href{https://arxiv.org/pdf/1401.4024.pdf}{[arXiv:1401.4024 [hep-th]]}.
		
\bibitem{Qiu:2011cy}
T.~Qiu, J.~Evslin, Y.~F.~Cai, M.~Li and X.~Zhang,
``Bouncing Galileon Cosmologies,''
\href{http://dx.doi.org/10.1088/1475-7516/2011/10/036}{JCAP {\bf 1110} (2011) 036}
\href{https://arxiv.org/pdf/1108.0593.pdf}{[arXiv:1108.0593 [hep-th]]}.

\bibitem{Easson:2011zy}
D.~A.~Easson, I.~Sawicki and A.~Vikman,
``G-Bounce,''
\href{http://dx.doi.org/10.1088/1475-7516/2011/11/021}{JCAP {\bf 1111} (2011) 021}
\href{https://arxiv.org/pdf/1109.1047.pdf}{[arXiv:1109.1047 [hep-th]]}.

\bibitem{Osipov:2013ssa}
M.~Osipov and V.~Rubakov,
``Galileon bounce after ekpyrotic contraction,''
\href{http://dx.doi.org/10.1088/1475-7516/2013/11/031}{JCAP {\bf 1311} (2013) 031}
\href{https://arxiv.org/pdf/1303.1221.pdf}{[arXiv:1303.1221 [hep-th]]}.

\bibitem{Qiu:2013eoa}
T.~Qiu, X.~Gao and E.~N.~Saridakis,
``Towards anisotropy-free and nonsingular bounce cosmology with scale-invariant perturbations,''
\href{http://dx.doi.org/10.1103/PhysRevD.88.043525}{Phys.\ Rev.\ D {\bf 88} (2013) no.4,  043525}
\href{https://arxiv.org/pdf/1303.2372.pdf}{[arXiv:1303.2372 [astro-ph.CO]]}.

\bibitem{Cai:2012va}
Y.~F.~Cai, D.~A.~Easson and R.~Brandenberger,
``Towards a Nonsingular Bouncing Cosmology,''
\href{http://dx.doi.org/10.1088/1475-7516/2012/08/020}{JCAP {\bf 1208} (2012) 020}
\href{https://arxiv.org/pdf/1206.2382.pdf}{[arXiv:1206.2382 [hep-th]]}.

\bibitem{Koehn:2013upa}
M.~Koehn, J.~L.~Lehners and B.~A.~Ovrut,
``Cosmological super-bounce,''
\href{http://dx.doi.org/10.1103/PhysRevD.90.025005}{Phys.\ Rev.\ D {\bf 90} (2014) no.2,  025005}
\href{https://arxiv.org/pdf/1310.7577.pdf}{[arXiv:1310.7577 [hep-th]]}.

\bibitem{Battarra:2014tga}
L.~Battarra, M.~Koehn, J.~L.~Lehners and B.~A.~Ovrut,
``Cosmological Perturbations Through a Non-Singular Ghost-Condensate/Galileon Bounce,''
\href{http://dx.doi.org/10.1088/1475-7516/2014/07/007}{JCAP {\bf 1407} (2014) 007}
\href{https://arxiv.org/pdf/1404.5067.pdf}{[arXiv:1404.5067 [hep-th]]}.

\bibitem{Qiu:2015nha}
T.~Qiu and Y.~T.~Wang,
``G-Bounce Inflation: Towards Nonsingular Inflation Cosmology with Galileon Field,''
\href{http://dx.doi.org/10.1007/JHEP04(2015)130}{JHEP {\bf 1504} (2015) 130}
\href{https://arxiv.org/pdf/1501.03568.pdf}{[arXiv:1501.03568 [astro-ph.CO]]}.

\bibitem{Kobayashi:2015gga}
T.~Kobayashi, M.~Yamaguchi and J.~Yokoyama,
``Galilean Creation of the Inflationary Universe,''
\href{http://dx.doi.org/10.1088/1475-7516/2015/07/017}{JCAP {\bf 1507} (2015) no.07,  017}
\href{https://arxiv.org/pdf/1504.05710.pdf}{[arXiv:1504.05710 [hep-th]]}.      

\bibitem{Wan:2015hya}
Y.~Wan, T.~Qiu, F.~P.~Huang, Y.~F.~Cai, H.~Li and X.~Zhang,
``Bounce Inflation Cosmology with Standard Model Higgs Boson,''
\href{http://dx.doi.org/10.1088/1475-7516/2015/12/019}{JCAP {\bf 1512} (2015) no.12,  019}
\href{https://arxiv.org/pdf/1509.08772.pdf}{[arXiv:1509.08772 [gr-qc]]}.

\bibitem{Ijjas:2016tpn}
A.~Ijjas and P.~J.~Steinhardt,
``Classically stable nonsingular cosmological bounces,''
\href{http://dx.doi.org/10.1103/PhysRevLett.117.121304}{Phys.\ Rev.\ Lett.\  {\bf 117} (2016) no.12,  121304}
\href{https://arxiv.org/pdf/1606.08880.pdf}{[arXiv:1606.08880 [gr-qc]]}.

\bibitem{Creminelli:2006xe}
P.~Creminelli, M.~A.~Luty, A.~Nicolis and L.~Senatore,
``Starting the Universe: Stable Violation of the Null Energy Condition and Non-standard Cosmologies,''
\href{http://dx.doi.org/10.1088/1126-6708/2006/12/080}{JHEP {\bf 0612} (2006) 080}
\href{https://arxiv.org/pdf/hep-th/0606090.pdf}{[hep-th/0606090]}.

\bibitem{Pirtskhalava:2014esa}
D.~Pirtskhalava, L.~Santoni, E.~Trincherini and P.~Uttayarat,
``Inflation from Minkowski Space,''
\href{http://dx.doi.org/10.1007/JHEP12(2014)151}{JHEP {\bf 1412} (2014) 151}
\href{https://arxiv.org/pdf/1410.0882.pdf}{[arXiv:1410.0882 [hep-th]]}.

\bibitem{Koehn:2015vvy}
M.~Koehn, J.~L.~Lehners and B.~Ovrut,
``Nonsingular bouncing cosmology: Consistency of the effective description,''
\href{http://dx.doi.org/10.1103/PhysRevD.93.103501}{Phys.\ Rev.\ D {\bf 93} (2016) no.10,  103501}
\href{https://arxiv.org/pdf/1512.03807.pdf}{[arXiv:1512.03807 [hep-th]]}.

\bibitem{Libanov:2016kfc}
M.~Libanov, S.~Mironov and V.~Rubakov,
``Generalized Galileons: instabilities of bouncing and Genesis cosmologies and modified Genesis,''
\href{http://dx.doi.org/10.1088/1475-7516/2016/08/037}{JCAP {\bf 1608} (2016) no.08,  037}
\href{https://arxiv.org/pdf/1605.05992v2.pdf}{[arXiv:1605.05992 [hep-th]]}.

\bibitem{Kolevatov:2016ppi}
R.~Kolevatov and S.~Mironov,
``Cosmological bounces and Lorentzian wormholes in Galileon theories with an extra scalar field,''
\href{http://dx.doi.org/10.1103/PhysRevD.94.123516}{Phys.\ Rev.\ D {\bf 94} (2016) no.12,  123516}
\href{https://arxiv.org/pdf/1607.04099v2.pdf}{[arXiv:1607.04099 [hep-th]]}.

\bibitem{Kobayashi:2016xpl}
T.~Kobayashi,
``Generic instabilities of nonsingular cosmologies in Horndeski theory: A no-go theorem,''
\href{http://dx.doi.org/10.1103/PhysRevD.94.043511}{Phys.\ Rev.\ D {\bf 94} (2016) no.4,  043511}
\href{https://arxiv.org/pdf/1606.05831.pdf}{[arXiv:1606.05831 [hep-th]]}.

\bibitem{Akama:2017jsa}
S.~Akama and T.~Kobayashi,
``Generalized multi-Galileons, covariantized new terms, and the no-go theorem for non-singular cosmologies,''
\href{http://dx.doi.org/10.1103/PhysRevD.95.064011}{Phys.\ Rev.\ D {\bf 95} (2017) no.6,  064011}
\href{https://arxiv.org/pdf/1701.02926.pdf}{[arXiv:1701.02926 [hep-th]]}.

\bibitem{Ijjas:2016vtq}
A.~Ijjas and P.~J.~Steinhardt,
``Fully stable cosmological solutions with a non-singular classical bounce,''
\href{http://dx.doi.org/10.1016/j.physletb.2016.11.047}{Phys.\ Lett.\ B {\bf 764} (2017) 289}
\href{https://arxiv.org/pdf/1609.01253.pdf}{[arXiv:1609.01253 [gr-qc]]}.  

\bibitem{Gleyzes:2014dya}
J.~Gleyzes, D.~Langlois, F.~Piazza and F.~Vernizzi,
``Healthy theories beyond Horndeski,''
\href{http://dx.doi.org/10.1103/PhysRevLett.114.211101}{Phys.\ Rev.\ Lett.\  {\bf 114} (2015) no.21,  211101}
\href{https://arxiv.org/pdf/1404.6495.pdf}{[arXiv:1404.6495 [hep-th]]}.

\bibitem{Cai:2016thi}
Y.~Cai, Y.~Wan, H.~G.~Li, T.~Qiu and Y.~S.~Piao,
``The Effective Field Theory of nonsingular cosmology,''
\href{http://dx.doi.org/10.1007/JHEP01(2017)090}{JHEP {\bf 1701} (2017) 090}
\href{https://arxiv.org/pdf/1610.03400v2.pdf}{[arXiv:1610.03400 [gr-qc]]}.		

\bibitem{Creminelli:2016zwa}
P.~Creminelli, D.~Pirtskhalava, L.~Santoni and E.~Trincherini,
``Stability of Geodesically Complete Cosmologies,''
\href{http://dx.doi.org/10.1088/1475-7516/2016/11/047}{JCAP {\bf 1611} (2016) no.11,  047}
\href{https://arxiv.org/pdf/1610.04207v2.pdf}{[arXiv:1610.04207 [hep-th]]}.

\bibitem{Cai:2017dyi} 
Y.~Cai and Y.~S.~Piao,
``A covariant Lagrangian for stable nonsingular bounce,''
\href{https://arxiv.org/pdf/1705.03401.pdf}{[arXiv:1705.03401 [gr-qc]]}.	
\end{thebibliography}
\end{document}